\documentclass[aps,twocolumn,showpacs,preprintnumbers,amsmath,amssymb,superscriptaddress]{revtex4}
\usepackage{graphicx,bm,mathrsfs,float}

\newcommand{\be}{\begin{equation}}
\newcommand{\ee}{\end{equation}}
\newcommand{\bea}{\begin{eqnarray}}
\newcommand{\eea}{\end{eqnarray}}
\newcommand{\bsube}{\begin{subequations}}
\newcommand{\esube}{\end{subequations}}
\newcommand{\Sec}[1]{Sec.\,\ref{#1}}
\newcommand{\Eq}[1]{Eq.\,(\ref{#1})}
\newcommand{\Eqs}[1]{Eqs.\,(\ref{#1})}
\newcommand{\Fig}[1]{Fig.\,\ref{#1}}
\newcommand{\Figs}[1]{Figs.\,\ref{#1}}
\newcommand{\rmS}{{\rm S}}
\newcommand{\rmB}{{\rm B}}
\newcommand{\rmT}{{\rm T}}

\newcommand{\rmi}{{\rm i}}

\newcommand{\rmD}{{\rm D}}

\newcommand{\ra}{\rangle}
\newcommand{\la}{\langle}
\newcommand{\lla}{\langle\!\langle}
\newcommand{\rra}{\rangle\!\rangle}

\newcommand{\omg}{\omega}
\newcommand{\Gam}{\Gamma}
\newcommand{\gam}{\gamma}

\newcommand{\lmd}{\lambda}
\newcommand{\vpl}{\varepsilon}
\newcommand{\epl}{\epsilon}

\newcommand{\bmq}{\bm{q}}

\newcommand{\tr}{{\rm tr}}

\begin{document}
\title{Anomalous current-electric field characteristics in transport through a nanoelectromechanical system}

\author{Chengjie Wu} 
\affiliation{Department of Physics, Zhejiang University of Science and Technology, Hangzhou 310023, China}

\author{Yi Ding}
\affiliation{Department of Physics, Zhejiang University of Science and Technology, Hangzhou 310023, China}
\affiliation{Department of Physics, Babe\c{s}-Bolyai University, 400084 Cluj-Napoca, Romania}

\author{Yiying Yan}
\affiliation{Department of Physics, Zhejiang University of Science and Technology, Hangzhou 310023, China}

\author{Yuguo Su}
\affiliation{Department of Physics, Zhejiang University of Science and Technology, Hangzhou 310023, China}

\author{Elijah Omollo Ayieta}
\affiliation{Department of Physics, University of Nairobi, Nairobi  30197, Kenya}

\author{Slobodan  Rado\v{s}evi\'{c}}
\affiliation{Department of Physics, Faculty of Science, University of Novi Sad, Trg Dositeja Obradovi\'{c}a 4, 21000 Novi Sad, Serbia}

\author{Georg  Engelhardt}
\affiliation{International Quantum Academy, Shenzhen 518048, China}
\affiliation{Shenzhen Institute for Quantum Science and Engineering, Southern University of Science and Technology, Shenzhen 518055, China}
\affiliation{Guangdong Provincial Key Laboratory of Quantum Science and Engineering, Southern University of Science and Technology, Shenzhen, 518055, China}

\author{Gernot Schaller}
\affiliation{Helmholtz-Zentrum Dresden-Rossendorf, Bautzner Landstra{\ss}e 400, 01328 Dresden, Germany}

\author{JunYan Luo}
\thanks{email: jyluo@zust.edu.cn}
\affiliation{Department of Physics, Zhejiang University of Science and Technology, Hangzhou 310023, China}

\date{\today}
	
\begin{abstract}
A thorough understanding of the correlation between electronic and mechanical degrees of freedom is crucial to the development of quantum devices in a nanoelectromechanical system (NEMS).
In this work, we first establish a fully quantum mechanical approach for transport through a NEMS device, which 
is valid for arbitrary bias voltages, temperatures, and electro-mechanical couplings.
We find an anomalous current-electric field characteristics at a low bias from, where the current decreases with a rising electric field. 
We reveal that this unique transport feature arises from a combined effect of mechanical motion and Coulomb blockade. At a large oscillation amplitude, the rapid increase of backward tunneling events, which is identified in both electron counting statistics and the phase space of charge-resolved Wigner functions, suppresses the forward current due to prohibition of double occupation.
In the opposite limit of strong damping, the oscillator dissipates its energy to the environment 
and relaxes to the ground state rapidly. 
Electrons then transport via the lowest vibrational state such that the net current and its corresponding noise have a vanishing dependence on the electric field.

\end{abstract}

\maketitle

\section{Introduction\label{thsec1}}

The rapid advancement in the state-of-the-art nanotechnology has made it possible to fabricate micromachines, where quantum mechanical effects have vital roles 
to play \cite{Met231559,Bil23274,Yan24783,You231697}. 
A nanoelectromechanical system (NEMS) represents such a notable quantum device where electronic transport becomes fundamentally coupled to nanomechanical vibrations \cite{She03R441,Gal07103201,Wei21724}. 
The exquisite sensitivity of electromechanical coupling has motivated groundbreaking applications ranging from quantum-limited detection of displacement \cite{Moz02018301,Moz04018303,Eta08785,Ble003845,Ane10061804}, 
charge \cite{Cle98160,Mee12115454,Zha15233505}, and spin \cite{Twa0663,Rug04329,Lam08136802}, to ultraprecise 
mass sensing \cite{Jen08533,Lei13154313,Cha12301,Zha18164503}.  Investigation of electron transport through a NEMS device is not only essential for a deeper understanding of the intricate interplay between electronic and vibrational degrees of freedom, but also significant for the future development of NEMS-based quantum devices.

Recent investigations of nonequilibrium transport through a NEMS device have revealed charge control capabilities at the single-electron level through synchronized mechanical oscillations \cite{Gor984526}.
Especially, a well ordered periodically shuttling behavior of electrons was 
predicted \cite{Isa98150,Wei9997,Nod02165312,McC03245415}.
A fully quantum mechanical treatment in the large bias limit (unidirectional transport) has unambiguously demonstrated the occurrence of a shuttling instability due to mechanical damping \cite{Don05237,Nov04248302,Nov03256801,Arm02035333}.
At finite bias, dynamics and transport characteristics have been analyzed within a semiclassical approach through phase-space approximation methods \cite{Wac19073009}. Especially, autonomous implementation of thermodynamic cycles \cite{Str21180605} and single electron emission source \cite{Wac22014011} have been theoretically demonstrated. 
A fully quantum-mechanical treatment at finite bias has so far been only discussed under the assumption of linear response of the motion of the charged oscillator to the electric field, effectively limiting their validity to a weak electromechanical coupling regime \cite{Lai15108501,Lai12175301}. It is therefore appealing to investigate the nonequilibrium transport through a NEMS device beyond the linear response assumption and reveal the influence of quantum uncertainty on transport characteristics for arbitrary bias and electromechanical couplings.

\begin{figure}
	\centering
	\includegraphics[width=8.5cm]{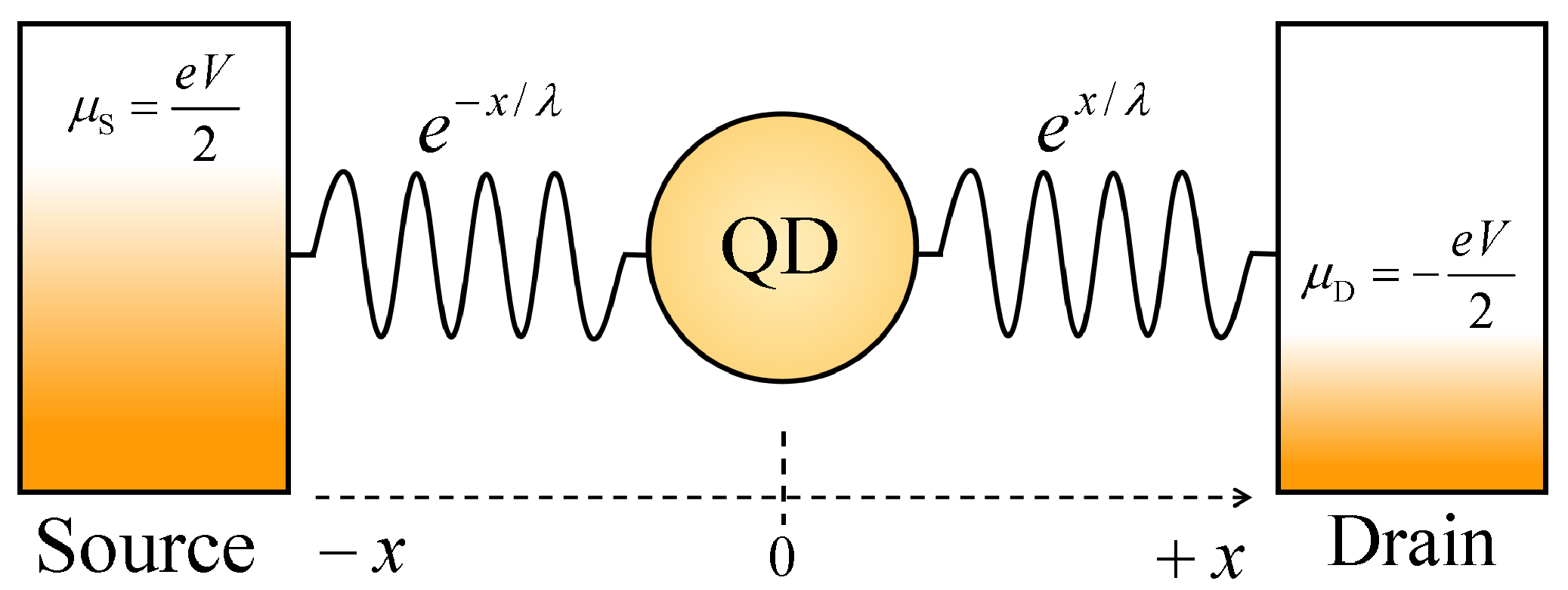}
	\caption{Schematics of an NEMS transport setup.	A QD oscillator is suspended between the source and the drain, with a bias voltage $eV=\mu_\rmS-\mu_\rmD$ applied across the two electrodes. Once an electron tunnels into the QD, it may vibrate between the two electrodes under the influence of the electric force. The entire system is embedded in a dissipative heat bath causing mechanical damping of the oscillator (not explicitly shown here).}
	\label{fig1}
\end{figure}

This work is dedicated to unveil the intriguing correlations between electronic and mechanical 
degrees of freedom with special attention paid to the effect of the electric field on the transport 
characteristics at low bias.
The transport NEMS device is schematically shown in \Fig{fig1}, where an oscillator is 
suspended between the source and the drain, with a finite bias voltage $eV=\mu_\rmS-\mu_\rmD$ 
applied across the two electrodes.
We first establish a fully quantum mechanical approach for transport through a NEMS device that 
is valid for arbitrary bias voltages and electro-mechanical couplings.
Once an electron jumps onto the oscillator from the source, the charged oscillator is subject to an 
electrostatic force $e{\cal E}$, where ${\cal E}=V/d$ is the effective electric field and $d$ is the effective distance between the two electrodes.
As for sufficiently large bias, electron loading happens dominantly at negative $x$ when the dot is closer to the source, the associated potential energy 
$\Delta V=-e {\cal E} x$ is positive, such that the charging process pumps energy into the NEMS.
For a given  bias regime $V$, a rise in ${\cal E}$ (by reducing $d$) means more electric energy is invested into the mechanical system and thus pushes the oscillator to higher excited states, which is expected to observable as an  increased current. 

However, we observe an anomalous behavior: The current does not necessarily increase with a rising electric field. This is ascribed to the backward tunneling of electrons at low bias.
Specifically, in the case of weak mechanical damping, an increase in the electric field excites the oscillator to higher vibrational modes with a large oscillation amplitude, which also leads to an enhancement of backward electron tunneling events, identified in both electron counting statistics and the phase space of charge-resolved Wigner functions.
In the strong Coulomb blockade regime that forbids double occupation, the rapid increase of backward tunneling events suppresses the forward current and eventually leads to a decreased net current. 
In the opposite limit of strong damping, the oscillator dissipates its energy to the environment and relaxes to the ground state very rapidly. 
Electrons are transported through the lowest vibrational state and the electric field has a vanishing influence. The net stationary current and the noise are almost independent of the electric field.

The rest of the paper is organized as follows. We start in \Sec{thsec2} with an
introduction of the NEMS device, followed in
\Sec{thsec3} by the establishment of the quantum master equation for the electron transport dynamics, which is applicable to arbitrary bias voltages and temperatures.
\Sec{thsec4} is devoted to a detailed investigation of the unique transport characteristics at 
low bias. Finally, we conclude our findings in \Sec{thsec5}.

\section{The Model System\label{thsec2}}

The transport NEMS setup is schematically shown in \Fig{fig1}, where a movable quantum dot (QD) is suspended between the source and drain electrodes. 
For small displacements from its equilibrium position, the center of mass
of the QD is well approximated to be bounded by a harmonic potential.
The oscillator is further immersed into a dissipative environment.
	The Hamiltonian of the entire system reads
	\begin{equation}
		\label{Htot}
		H_\rmT=H_\rmS+H_{\rm el}+H_{\rm tun}+H_\rmB+H_{\rm int},
	\end{equation}
	where the first term
	\begin{equation}
		\label{HS}
		H_\rmS=\frac{p^2}{2 m}+\frac{1}{2}m \omg_0^2 x^2+(\vpl_0-e{\cal E}x)c_0^\dag c_0
	\end{equation}
	describes the reduced system: A single level ($\vpl_0$) QD which is modeled as a simple harmonic oscillator of mass $m$ and natural frequency $\omega_0$, in which $c_0^\dag$ ($c_0$) is the creation (annihilation) operator for an electron in the QD.
	The bias voltage ($V$) between the source and the drain generates an electric field ${\cal E}=V/d$ (with $d$ the effective distance between the two electrodes), giving
	rise to an electric influence on the mechanical dynamics.
	The small size of the QD implies a very small capacitance and thus a 
	very large charging energy which is assumed to be the largest energy scale in this work. 
	That means at most one excess electron can occupy the QD (Coulomb blockade)
	and we can thus describe the electronic state of the QD as a two-state system: $|g\ra$-empty and $|e\ra$-charged by an excess electron.

	The second term $H_{\rm el}$ describes the source and drain electrodes, which are modeled as reservoirs of noninteracting electrons
	\begin{equation}
		H_{\rm el}=\sum_{\nu=\rmS,\rmD}\sum_k \epl_{\nu k } c_{\nu k}^\dag c_{\nu k},
	\end{equation}
	where $c_{\nu k}^\dag$ $(c_{\nu k})$ is the creation (annihilation) operator for an electron with momentum $k$ in the source ($\nu=\rmS$) or the drain ($\nu=\rmD$).
	Each electrode is assumed to be in equilibrium, so that they can be characterized by the
	Fermi distributions $f_{\nu}(\omega) = \{1 + e^{\beta_\nu (\omega - \mu_\nu)}\}^{-1}$, with the inverse
	temperature $\beta_\nu = (k_{\rm B}T_\nu)^{-1}$ and the chemical potential $\mu_\nu$  in the 
	electrode $\nu$ ($\nu=\rmS$ or $\rmD$).
	The bias across the device is thus given by $ e V = (\mu_\rmS-\mu_\rmD)$.

Electrons can tunnel between the electrodes and the QD with tunneling
amplitudes which are assumed to be  exponentially dependent on the position of the movable QD within the range where the harmonic approximation to the oscillator holds.
This is due to the exponentially decreasing/increasing overlapping of the electronic wave functions.
The tunnel coupling Hamiltonian described by the third term of \Eq{Htot} thus can be expressed as 
\begin{equation}\label{Htun}
H_{\rm tun}=\sum_{\nu=\rmS,\rmD}\sum_{k} (t_{\nu k} e^{\frac{x}{\lambda_\nu}}c^\dag_{\nu k} c_0+{\rm h.c.}),
\end{equation}
where the tunneling amplitudes $t_{\nu k} e^{\frac{x}{\lmd_\nu}}$ explicitly depend on the position $x$, with $\lmd_{\rmS/\rmD}=\mp \lambda$ and $\lambda$ the tunneling length.
	The bare tunneling width between the oscillator and the electrode $\nu$ ($\nu\in\{\rmS,\rmD\}$) is given by $\Gamma_{\nu}(\omega)=2\pi\sum_k |t_{\nu k}|^2 \delta(\omega-\vpl_{\nu k})$.
	In what follows, we assume the wide-band limit such that the tunneling width is frequency independent, i.e.,
	$\Gamma_{\nu}(\omega)=\Gamma_{\nu}$.

	
The QD oscillator is further subject to a dissipative environment that is modeled as a collection of phonons
coupled to the oscillator by a weak bilinear interaction as following \cite{Gar99}:
\bsube
\begin{gather}
	H_\rmB=\sum_{\bmq}  \omg_{\bmq} a_{\bmq}^\dag a_{\bmq},
	\\
	H_{\rm int}=\sum_{\bmq} g_{\bmq} \sqrt{\frac{2m\omg_0}{\hbar}}x(a_{\bmq}^\dag+ a_{\bmq}),
\end{gather}
\esube
where $a_{\bmq}^\dag$ ($a_{\bmq}$) is the creation (annihilation) operator of a bath phonon with
wave number $\bmq$ and $g_{\bmq}$ represents the coupling strength. The corresponding damping rate is given by $\gam(\omg_0)=2\pi\sum_{\bmq} |g_{\bmq}|^2 \delta(\omg-\omg_0)$.

In order to analyze different time scales associated with the resonator oscillations, the 
tunneling events, as well as the mechanical coupling, it is instructive to eliminate the coupling term of the oscillator and the charge by performing a  Lang-Firsov 
transform \cite{Mah90,Lan621301}
\be
{\cal S}=\exp\left\{i\frac{e{\cal E}p}{\hbar m\omega^2_0}c_0^\dag c_0\right\},
\ee
which satisfies ${\cal S}{\cal S}^\dag={\cal S}^\dag {\cal S}=1$. This unitary transformation gives a displacement of the oscillator conditioned on 
the electronic occupation in the quantum dot. The Liouvillian equation of the entire system for the displaced 
state $\varrho_{\rm T}={\cal S} \rho_{\rm T} {\cal S}^\dag$ reads
\be
\dot{\varrho}_{\rm T}=-i[{\mathbf H}_{\rm T},\varrho_{\rm T}],
\ee
where ${\mathbf H}_{\rm T}={\cal S} H_{\rm T} {\cal S}^\dag$ is the total Hamiltonian in the 
displaced picture. 
The reduced system Hamiltonian becomes 
\bea
{\mathbf H}_{\rm S}
&=&\frac{p^2}{2 m}+\frac{1}{2}m \omg_0^2 x^2+\tilde{\vpl}_0 c_0^\dag c_0
\nonumber \\
&=&\hbar\omega_0 \left(a^\dag a+\frac{1}{2}\right)+\tilde{\vpl}_0 c_0^\dag c_0,
\eea
where the mechanical and electronic degrees of freedom are decoupled, with $a^\dag$ ($a$)
the creation (annihilation) operators for oscillator excitations and
$\tilde{\vpl}_0=\vpl_0-e^2{\cal E}^2/m\omega_0^2$.
Although in the polaron frame the QD energy alone decreases with the field strength, the tunneling Hamiltonian in this frame becomes
\be\label{htun}
{\mathbf H}_{\rm tun}=\sum_{\nu=\rmS,\rmD}\sum_k t_{\nu k} F_{\nu} c^\dag_{\nu k} c_0+{\rm h.c.},
\ee
where, for convenience, we have introduced the short-hand notation $F_{\nu}=e^{\frac{x}{\lmd_\nu}}e^{-i\frac{e{\cal E}p}{\hbar m\omega^2_0}}$, 
such that charging goes along with the creation of multiple vibrational quanta.
The Hamiltonian for the source and drain is not affected, i.e., ${\mathbf H}_{\rm el}=H_{\rm el}$.
In the following we set $\hbar=e=k_\rmB=1$, unless stated otherwise.

\section{\label{thsec3}Electron-Number-Resolved Quantum Master Equation}
\subsection{\label{SRQME}The quantum master equation}

In the displaced picture, the electronic and mechanical dynamics of the QD oscillator is described by a quantum master equation (QME) for the
corresponding reduced density matrix $\varrho(t)={\rm tr}_{\rm B}[\varrho_{\rm T}(t)]$, with ${\rm tr}_{\rm B}[\cdots]$ denoting the trace over the electrodes and heat bath degrees of freedom. It can be derived from the Liouville-von Neumann equation 
for the total system by projecting out the degrees of freedom of the electrodes and the thermal bath under the second-order perturbation expansion.
It is then transformed  from the displaced picture back to the original picture via $\rho(t)={\cal S}^\dag \varrho(t) {\cal S}$.
Furthermore, to characterize the electron transport properties, we introduce a counting field $\chi$ associated with the numbers of transferred electrons \cite{Esp091665,Bag03085316}. One finally
arrives at a counting-field-dependent QME for the $\chi$-dependent density matrix $\rho(\chi,t)$
\begin{align}\label{QME}
	\dot{\rho}(\chi,t)
	&={\cal W}(\chi)\rho(\chi,t)
	\nonumber \\
	&=\{{\cal L}_{\rm coh}+{\cal L}_{\rm tun}(\chi)+{\cal L}_{\rm damp}\}\rho(\chi,t),
\end{align}
where the first term
\be
{\cal L}_{\rm coh}\rho(\chi,t)=-\rmi [H_{\rm S},\rho(\chi,t)]
\ee
describes the coherent evolution of the QD oscillator in \Eq{HS}.
The second term accounts for the tunneling between the electrodes and the QD. Under the second-order Born-Markov approximation in terms of the
system-electrode coupling, this term is given by (a detailed derivation is referred to Appendix \ref{thAppA})
\begin{widetext}
\begin{align}\label{QME1}
{\cal L}_{\rm tun}(\chi)\rho(\chi,t)=
&
-\frac{\Gamma_\rmS}{2}
		\left[ e^{-\frac{x}{\lambda}} c_0, \Upsilon_{\rmS}^{(+)}({L}_{\rm osc})\rho(t)-\rho(t) 
		\Upsilon_{\rmS}^{(-)}({L}_{\rm osc})\right]		
		-\frac{\Gamma_\rmS}{2}
		\left[c_0^\dag e^{-\frac{x}{\lambda}}, \Upsilon_{\rmS}^{(-)}(-{L}_{\rm osc})\rho(t)-\rho(t) 
		\Upsilon_{\rmS}^{(+)}(-{L}_{\rm osc})\right]
		\nonumber \\
		&-\frac{\Gamma_\rmD}{2}\left[
		e^{\frac{x}{\lambda}} c_0\Upsilon_{\rmD}^{(+)}({L}_{\rm osc})\rho(t)+\rho(t) 
		\Upsilon_{\rmD}^{(+)}(-{L}_{\rm osc})\,c_0^\dag e^{\frac{x}{\lambda}}
		+
		c_0^\dag\, e^{\frac{x}{\lambda}} \Upsilon_{\rmD}^{(-)}(-{L}_{\rm osc})\rho(t)+\rho(t) 
		\Upsilon_{\rmD}^{(-)}({L}_{\rm osc})e^{\frac{x}{\lambda}} c_0\right]
		\nonumber \\
		&+\frac{\Gamma_\rmD}{2}\left[
		e^{\frac{x}{\lambda}} c_0\rho(t)\Upsilon_{\rmD}^{(-)}({L}_{\rm S})+\Upsilon_{\rmD}^{(-)}(-{L}_{\rm S})\rho(t) c_0^\dag e^{\frac{x}{\lambda}}
		\right] e^{-i\chi}
		\nonumber \\
		&+\frac{\Gamma_\rmD}{2}
		\left[c_0^\dag e^{\frac{x}{\lambda}}\rho(t)\Upsilon_{\rmD}^{(+)}(-{L}_{\rm S})+
		\Upsilon_{\rmD}^{(+)}({L}_{\rm S})\rho(t) e^{\frac{x}{\lambda}}c_0 \right] e^{i\chi},
	\end{align}
\end{widetext}
where the terms with $\Gamma_\rmS$ describe the tunneling of electrons between the source and the oscillator, and the term with $\Gamma_\rmD$ denotes those between the oscillator and the drain.
The counting field $\chi$ is introduced to count the electrons having tunneled  from the drain into the QD ($\chi$) or from the QD to the drain
($-\chi$).
The counting statistics of transferred electrons between the source and the oscillator can also be accounted for in an analogous way if one introduces a corresponding counting field.
Here, for compactness, we have introduced the following shorthand notations
\bsube\label{UpsilonPM0}
\begin{align}
	&\Upsilon_{\nu}^{(\pm)}(+{L}_{\rm osc})
	={\cal S}^\dag f_{\nu}^{(\pm)}(\tilde{\varepsilon}_0+{L}_{\rm osc})[c_0^\dag\! F_{\nu}^\dag]{\cal S},\label{UpsilonPM1}
	\\
	&\Upsilon_{\nu}^{(\pm)}(-{L}_{\rm osc})
	={\cal S}^\dag f_{\nu}^{(\pm)}(\tilde{\varepsilon}_0-{L}_{\rm osc})[F_{\nu}c_0]{\cal S},\label{UpsilonPM2}
\end{align}
\esube
where ${L}_{\rm osc}$ is the superoperator associated with the free oscillator Hamiltonian, 
i.e., ${L}_{\rm osc}[\cdots]=[H_{\rm osc},(\cdots)]$, with $H_{\rm osc}=\frac{p^2}{2 m}+\frac{1}{2}m \omg^2 x^2$ 
the bare Hamiltonian of the oscillator; $f_\nu^{(+)}(\omega)=f_\nu(\omega)$ is the usual Fermi function in the source ($\nu={\rm S}$) or drain ($\nu={\rm D}$), and $f_\nu^{(-)}(\omega)\equiv1-f_\nu^{(+)}(\omega)$.
The application of ${\cal S}^\dag (\cdots){\cal S}$ is used to transform the master equation from the displaced picture back to the original basis. 
A detailed evaluation of \Eqs{QME1} and (\ref{UpsilonPM0}) is given in
 Appendix \ref{thAppA}.  
It is worthwhile to mention that in the limit of large bias ($V\to\infty$), the Fermi functions $f_{\nu}^{(\pm)}(\tilde{\varepsilon}_0\pm{L}_{\rm osc})$ in \Eqs{UpsilonPM1} and (\ref{UpsilonPM2}) can be well approximated
by either 1 or 0, such that the QME (\ref{QME}) reduces to the one widely used in the literature for unidirectional transport \cite{Nov03256801}.  
In the limit when the harmonic oscillation frequency is much smaller than the coupling rates to the source, drain and phonon baths, the quantum master equation (\ref{QME}) reduces to a classical rate equation which works at finite temperature and bias \cite{Str21180605,Wac19024001,Wac19073009} (see also our derivation in Appendix \ref{thAppB}). 
Our work thus provides a unifying and compact formalism to account for transport through NEMS systems for arbitrary oscillator frequency, bias and electro-mechanical coupling quantum mechanically.

The last term in \Eq{QME} accounts for the interaction of the oscillator with the heat bath, leading to 
mechanical damping of the oscillator. In second order perturbation, it is readily given by \cite{Koh975236}
\be\label{caldamp}
{\cal L}_{\rm damp}\rho(t)=-i\frac{\gamma}{2} [x,\{p, \rho\}]-\gam m \omg_0 (\bar{N}+{\textstyle \frac{1}{2}})[x,[x,\rho]],
\ee
where $\bar{N}=[e^{\beta_{\rm B}\omg_0}-1]^{-1}$ is the Bose-Einstein distribution at oscillator frequency $\omg_0$ and inverse bath temperature $\beta_{\rm B}=(k_{\rm B}T_{\rm B})^{-1}$.

Specifically, let us consider the reduced dynamics in the electronic subspaces: $|0\ra$ and $|1\ra$, corresponding an empty and occupied oscillator, respectively. 
The corresponding $\chi$-resolved QME in \Eq{QME} for the reduced density matrices $\rho_{00}(t)=\la 0|\rho(t)|0\ra$ and $\rho_{11}(t)=\la 1|\rho(t)|1\ra$ is given by (The off-diagonal reduced density matrices are dynamically decoupled from the diagonal ones and decay to zero in the steady state, and thus are irrelevant)
\begin{widetext}
\bsube\label{QME4}
\begin{align}
\dot{\rho}_{00}=&-\rmi[H_{\rm osc},\rho_{00}]
-\frac{1}{2}\sum_{\nu} \Gamma_{\nu}
		\left\{e^{\frac{x}{\lambda_\nu}} {\cal S}^\dag_{11}  f_{\nu}^{(+)}(\tilde{\varepsilon}_0+{L}_{\rm osc})[F^\dag_{\nu}]\rho_{00} 
		+\rho_{00} f_{\nu}^{(+)}(\tilde{\varepsilon}_0-{L}_{\rm osc})[F_{\nu}] {\cal S}_{11} e^{\frac{x}{\lambda_\nu}}\right\}
		\nonumber \\
		&+\frac{1}{2}\sum_{\nu} \Gamma^{(-i\chi)}_{{\nu}}
		\left\{
		e^{\frac{x}{\lambda_\nu}} \rho_{11} {\cal S}^\dag_{11} f_{\nu}^{(-)}(\tilde{\varepsilon}_0+{L}_{\rm osc})[F^\dag_{\nu}]
		+f_{\nu}^{(-)}(\tilde{\varepsilon}_0-{L}_{\rm osc})[F_{\nu}] {\cal S}_{11} \rho_{11} e^{\frac{x}{\lambda_\nu}}\right\}+{\cal L}_{\rm damp}\rho_{00},\label{rhodsa}
		\\
		\dot{\rho}_{11}=&-\rmi[H_{\rm osc}-{\cal E}x,\rho_{11}]
		+\frac{1}{2}\sum_{\nu} \Gamma_{\nu}^{(i\chi)}
		\left\{e^{\frac{x}{\lambda_\nu}} \rho_{00}f_{\nu}^{(+)}(\tilde{\varepsilon}_0-{L}_{\rm osc})[F_{\nu}] {\cal S}_{11}
		+ {\cal S}_{11}^\dag f_{\nu}^{(+)}(\tilde{\varepsilon}_0+{L}_{\rm osc})[F^\dag_{\nu}] \rho_{00}  e^{\frac{x}{\lambda_\nu}}\right\}
		\nonumber \\
		&-\frac{1}{2}\sum_{\nu} \Gamma_{\nu}
		\left\{e^{\frac{x}{\lambda_\nu}}  f_{\nu}^{(-)}(\tilde{\varepsilon}_0-{L}_{\rm osc})[F_{\nu}]{\cal S}_{11} \rho_{11}
		+\rho_{11}  {\cal S}^\dag_{11}  f_{\nu}^{(-)}(\tilde{\varepsilon}_0+{L}_{\rm osc})[F^\dag_{\nu}] e^{\frac{x}{\lambda_\nu}}\right\}
		+{\cal L}_{\rm damp}\rho_{11},
	\end{align}
	\esube
\end{widetext}
where, for compactness, we have introduced $\Gamma^{(\pm i\chi)}_{{\rmD}}=\Gamma_{{\rmD}} e^{\pm i\chi}$ 
and $\Gamma^{(\pm i\chi)}_{{\rmS}}=\Gamma_{{\rmS}}$, since we only count electron transport between the oscillator and the drain, and ${\cal S}_{11}=\la e |{\cal S}|e\ra=e^{i\frac{{\cal E}p}{m\omega^2_0}}$.
Note that each density matrix element $\rho_{00}$ or $\rho_{11}$ is still a matrix in the Hilbert space of the oscillator.
Equation (\ref{QME4}) provides a convenient starting point for evaluating various transport characteristics based on the FCS.

\subsection{Electron counting statistics}
The cumulant-generating function of the electron counting statistics is simply given by \cite{Esp091665}
\be
e^{{\cal F}(\chi,t)}=\tr [\rho(\chi,t)],
\ee
where $\rho(\chi,t)$ is the solution of \Eq{QME}. In the stationary limit ($t\to\infty$), it has been revealed that
the cumulant-generating function is determined by \cite{Bag03085316,Gro06125315,Fli05475,Kie06033312} 
\be
{\cal F}(\chi,t)\to \Lambda_0(\chi) t,
\ee
where $\Lambda_0(\chi)$ is the unique eigenvalue that solves the eigenvalue problem for the total
Liouvilian ${\cal W}({\chi})$ in \Eq{QME}
\be\label{calW}
{\cal W}(\chi)|0(\chi)\rra=\Lambda_0(\chi)|0(\chi)\rra
\ee
and satisfies $\Lambda_0(\chi\to0)=0$. Here $|0(\chi)\rra$ is the right eigenvectors corresponding to $\Lambda_0(\chi)$.
In the limit $\chi\to0$, it reduces to the stationary solution of the quantum master equation (\ref{QME}): $|0\rra\equiv|0(\chi=0)\rra=\rho^{\rm st}$.
One may find this unique eigenvalue $\Lambda_0(\chi)$  in a spirit similar to the Rayleigh-Schr{{\"{o}}}dinger
perturbation theory \cite{Fli10155407}.
To this end, we split the total Liouvilian ${\cal W}({\chi})$ in \Eq{QME} into two parts
\be
{\cal W}({\chi})={\cal W}_0+\overline{{\cal W}}({\chi}),
\ee
where ${\cal W}_0\equiv{\cal W}({\chi}=0)$ is the `unperturbed' part that solves for the stationary solution ${\cal W}_0|0\rra=0$.
It corresponds to the zeroth order solution of \Eq{calW} such that $\Lambda_0(\chi)$ can be considered as evolved adiabatically from $\Lambda_0(\chi=0)=0$.
Accordingly, one can introduce the left eigenvector $\lla \tilde{0}|$ that satisfies $\lla \tilde{0}|{\cal W}_0=0$, which
corresponds to $\lla \tilde{0}|=$\{vec(\textbf{I}),vec(\textbf{I})\} (a row vector 
obtained by concatenating the two flattened identity matrices with the same dimension that is used to approximately represent the oscillator) 
such that $\lla \tilde{0}|0\rra=1$ implements the trace of the stationary state.

This further motivates us to introduce the projection operators ${\cal P}=|0\rra\lla \tilde{0}|$ and ${\cal Q}=1-{\cal P}$ satisfying ${\cal P}^2={\cal P}$, ${\cal Q}^2={\cal Q}$, and ${\cal P}{\cal Q}={\cal Q}{\cal P}=0$.
With these definitions, the perturbation expansion can be carried out in terms of the perturbation operator $\overline{{\cal W}}(\chi)\equiv{\cal W}(\chi)-{\cal W}_0$.
Applying the left eigenvector $\lla \tilde{0}|$ on both sides of the eigenvalue equation (\ref{calW}), one arrives at
\be\label{lmd0}
\Lambda_0(\chi)=\lla \tilde{0}| \overline{{\cal W}}(\chi) |0(\chi)\rra,
\ee
where we have used $\lla \tilde{0}|{\cal W}_0=0$ and the conventional normalization $\lla \tilde{0}|0(\chi)\rra =1$.
It is readily found that $\Lambda_0(\chi)$ obeys the equation \cite{Fli10155407}
\be\label{Lamdfinal}
\Lambda_0(\chi)=\lla \tilde{0}| \overline{{\cal W}}(\chi) |\sum_{n=0}^\infty
\left\{{\cal R}[\Lambda_0(\chi)-\overline{{\cal W}}(\chi)]\right\}^n|0\rra,
\ee
where  ${\cal R}$ is a pseudoinverse operator ${\cal R}={\cal Q}{\cal W}^{-1}_0{\cal Q}$.
Equation (\ref{Lamdfinal}) serves as a convenient starting point to evaluate the cumulants.
Taylor expanding all the quantities around $\chi=0$ as $\Lambda_0(\chi)=\sum_{k=1}^{\infty}\frac{(i\chi)^k}{k!} \lla I^k\rra$ and
$\overline{{\cal W}}(\chi)=\sum_{k=1}^{\infty}\frac{(i\chi)^k}{k!}\overline{{\cal W}}^{(k)}$,
a recursive scheme can be established to evaluate the cumulants up to an arbitrary 
order, in principle.
For instance, the first two cumulants are given by
\begin{subequations}
\begin{gather}
		\lla I \rra=\lla \tilde{0}| \overline{{\cal W}}^{(1)}|0\rra,\label{c1}
		\\
		\lla I^2 \rra=\lla \tilde{0}| [\overline{{\cal W}}^{(2)}-2\overline{{\cal W}}^{(1)} {\cal R}\overline{{\cal W}}^{(1)}]|0\rra.\label{c2}
\end{gather}
\end{subequations}
Higher-order cumulants can be obtained in a recursive manner.

\section{RESULTS AND DISCUSSION\label{thsec4}}

\begin{figure}
	\centering
	\includegraphics[width=8.5cm]{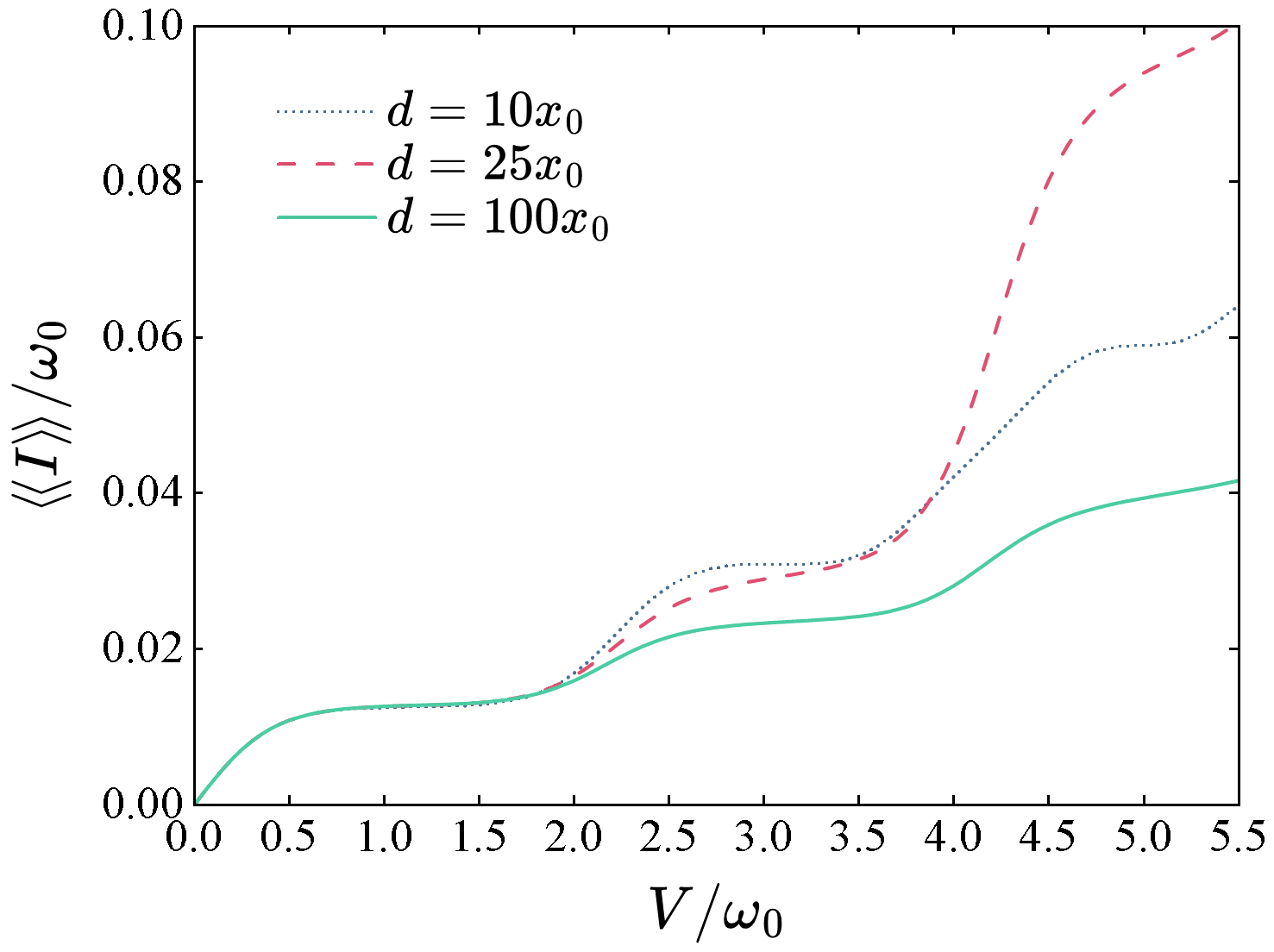}
	\caption{The $I-V$ characteristics for three different electric fields (effective distances $d=10x_0$, $d=25x_0$, and $d=100x_0$) where $x_0=1/\sqrt{2m\omega_0}$ is the zero-point uncertainty. The other plotting parameters are $\Gamma_{\rm S}=\Gamma_{\rm D}=0.01\omega_0$, $\gamma=0.02\omega_0$, $\lambda=2x_0$, and electrode temperature $\beta_{\rm S}=\beta_{\rm D}=10\omega^{-1}_0$. The heat bath temperature is 0.}
	\label{fig2}
\end{figure}

It has been observed in experiment that the steady-state current through an oscillator increases with bias in a ``step-like'' manner \cite{Par0057}. This is due to the fact that whenever one more vibration energy level enters the bias window defined by the chemical potentials of the source and the drain, one more channel opens up for transport.
In the limit of a large bias, it is predicted that the current reaches its maximum value, corresponding to the shuttling of one electron per each cycle ($\lla I\rra/\omega_0 =2\pi$)  for a weak mechanical damping \cite{Nov03256801}.
In the low bias regime, the effective electric field becomes an important quantity that can influence nanoelectromechanical transport characteristics. 
Once an electron jumps onto the oscillator from the source, the charged oscillator is subject to an 
electrostatic force $e{\cal E}$, where ${\cal E}=V/d$ is the effective electric field.
Energy is then pumped into the mechanical system.
For a given  bias $V$, a rise in ${\cal E}$ (by reducing $d$) means more electric energy is invested into the mechanical system and thus pushes the oscillator to higher excited states such that one expects to observe an increased current.

To illustrate this, we plot in Fig. \ref{fig2} the $I-V$ characteristics for various electric fields (via changing the effective distance $d$) with a weak damping. 
We observe that the current in general increases with bias in a ``step-like'' way, in agreement with the observations in experiments \cite{Par0057}. 
At $V=\omega_0$, only the ground-state mechanical level is available such that one observes equivalent currents under different electric fields.

\begin{figure}
	\centering
	\includegraphics[width=8.5cm]{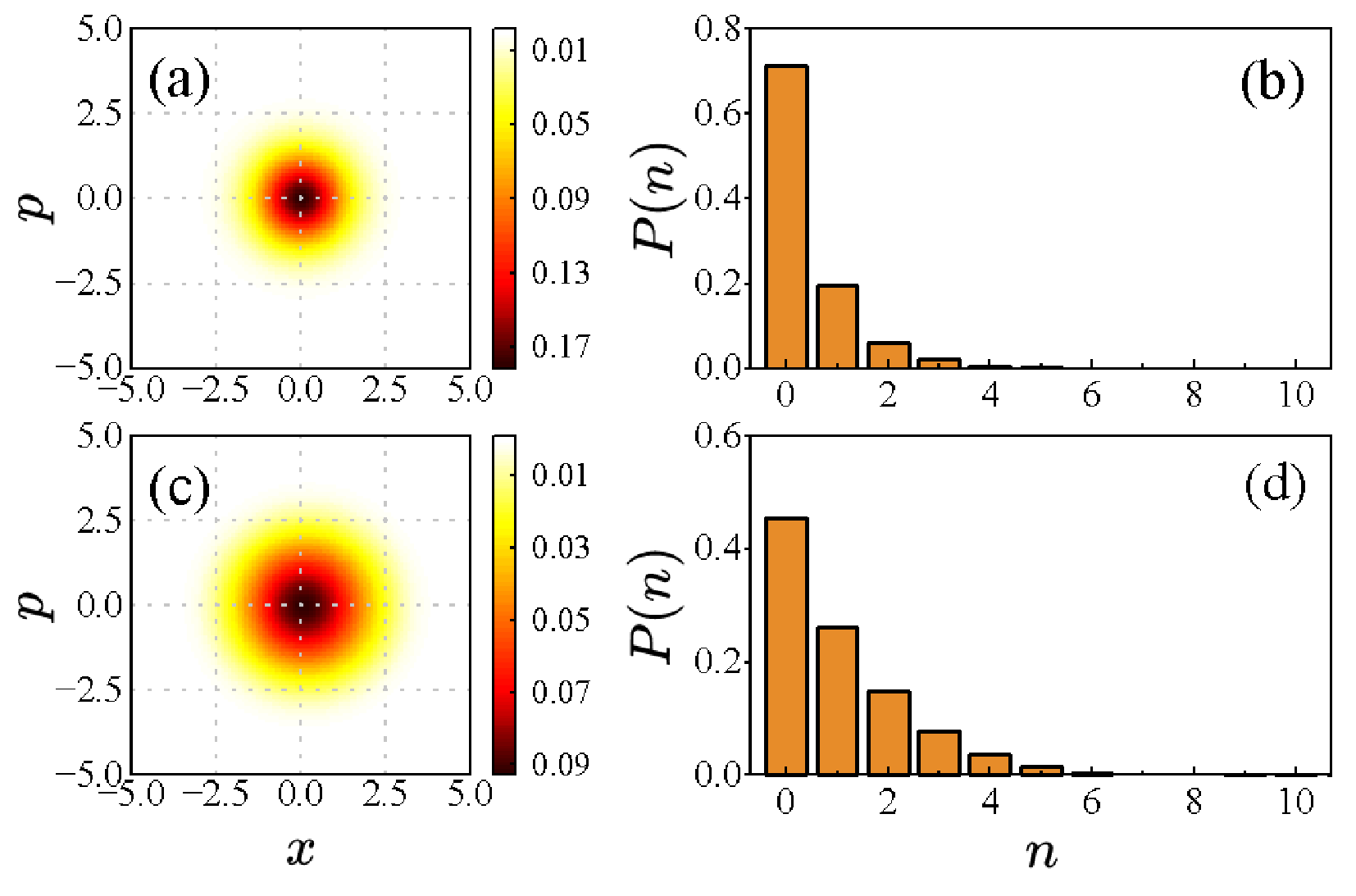}
	\caption{\label{fig3} (a) The Wigner function and (b) the phonon occupation probability distribution of the oscillator under the bias $V=3\omega_0$ for a weak electric field corresponding to $d=100x_0$. (c) The Wigner function and (d) the phonon occupation probability distribution at a strong electric field corresponding to $d=10x_0$. The other plotting parameters are the same as those in \Fig{fig2}.}
\end{figure}

As the bias increases, more vibration modes are activated and the current rises to the second plateau, where the electric field starts to 
influence the current, see for instance, different currents at $V=3\omega_0$ in \Fig{fig2}.
For a small electric field ($d=100x_0$),  the lowest two vibrational modes (ground and first excited states) dominate, which is indicated by  the phonon occupation probabilities of the oscillator 
$P(n)=\la n|(\rho^{\rm st}_{00}+\rho^{\rm st}_{11})|n\ra$, as shown in \Fig{fig3}(b).
In the phase space of the oscillator's quantum state represented by the Wigner 
function \cite{Koh975236,Iot17115420} $W(x,p)=W_{00}(x,p)+W_{11}(x,p)$, with the charge-resolved components
\begin{equation}
W_{00/11}(x,p)=\int\limits_{-\infty}^{\infty} \dfrac{d\xi}{2\pi}\bigg\langle x-\dfrac{\xi}{2}\bigg|\rho_{00/11}^{\rm st}\bigg|x+\dfrac{\xi}{2}\bigg\rangle e^{ip\xi},
\end{equation}
it simply shows a blurred spot at the center, see \Fig{fig3}(a).
As the electric field increases, more vibrational modes are involved, which are indicated in \Fig{fig3}(d) for $d=10x_0$. The corresponding Wigner distribution shows an enlarged spot at the center. 
As a result, one observes a larger current in comparison with that for a weak electric field.
So far, the results are consistent with our naive expectation that an increased electric field should normally lead to an enhanced current at 
a given bias.

\begin{figure}
	\centering
	\includegraphics[width=8cm]{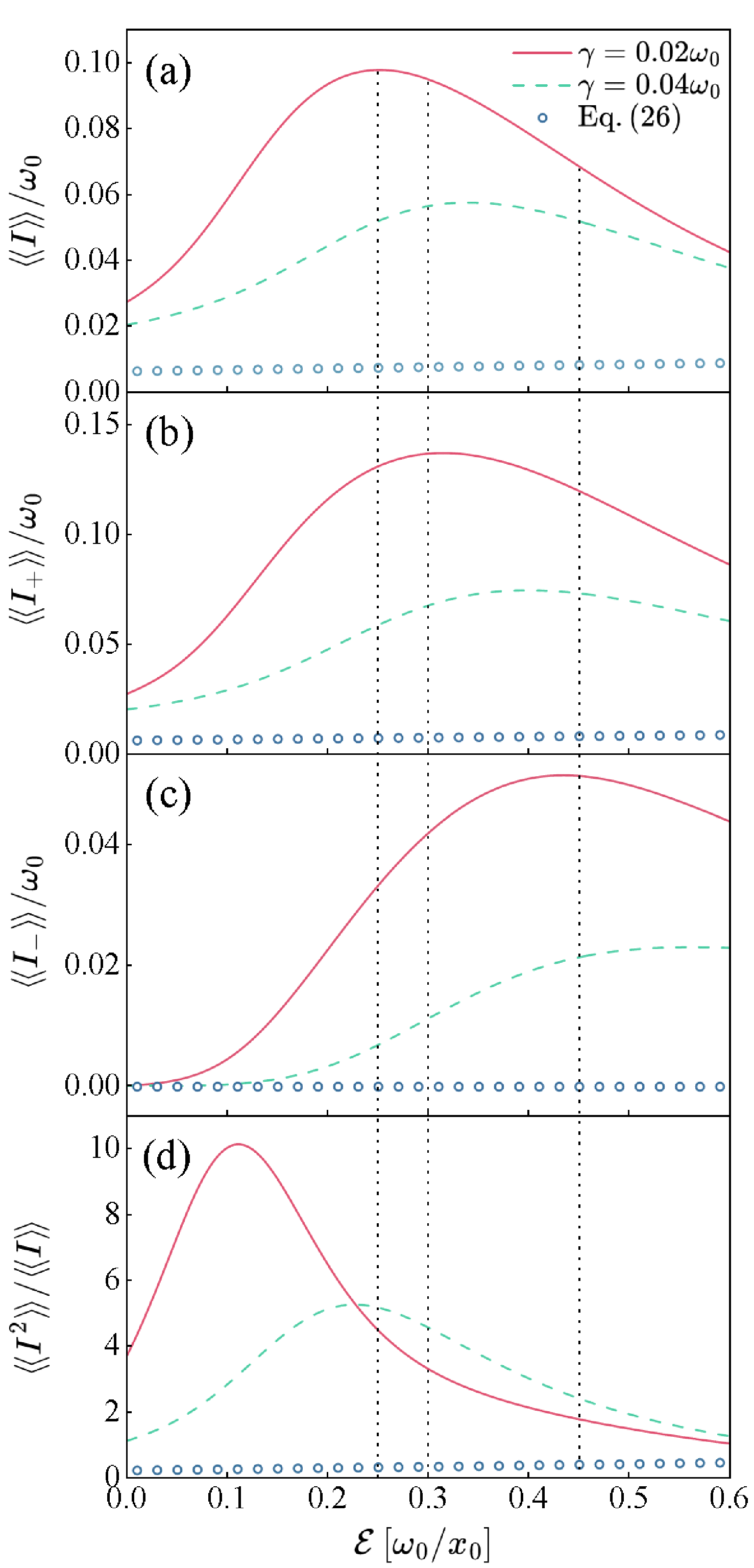}
	\caption{\label{fig4}The stationary current ($\lla I\rra$), forward ($\lla I_+\rra$) and backward ($\lla I_-\rra$) components, and the Fano factor ($\lla I^2\rra/\lla I\rra$) versus electric field under a given bias $V=5\omega_0$ 
		for various mechanical dampings. The other plotting parameters are the same as those in \Fig{fig2}.}
\end{figure}

Strikingly, as the bias voltage increase to $V=5\omega_0$, one finds a stronger electric field does not necessarily give rise to a larger current, as shown by the dashed and dotted curves in \Fig{fig2}, where  a stronger electric field ($d=10x_0$) results in a smaller current than that of a medium electric field ($d=25x_0$). That means the current does not increase monotonically with the electric field.
For a detailed analysis, we unravel the stationary current into the forward and backward current contributions as
\bsube
\begin{align}
	\lla I\rra=&\lla I_+\rra-\lla I_-\rra,\label{I}
\end{align}
where
\begin{align}
	\lla I_+\rra =&\frac{\Gamma_{\rm D}}{2}{\rm Tr_{osc}}\Big\{e^{x/\lambda} \rho_{11} {\cal S}^\dag_{11} f_{\rm D}^{(-)}(\tilde{\varepsilon}_0+{L}_{\rm osc})[F^\dag_{\rm D}]\notag\\&+f_{\rm D}^{(-)}(\tilde{\varepsilon}_0-{L}_{\rm osc})[F_{\rm D}] {\cal S}_{11} \rho_{11} e^{x/\lambda}	\Big\},\label{I1}
\end{align}
and
\begin{align}
	\lla I_-\rra =&\frac{\Gamma_{\rm D}}{2}{\rm Tr_{osc}}\Big\{e^{x/\lambda}  \rho_{00}  f_{\rm D}^{(+)}(\tilde{\varepsilon}_0-{L}_{\rm osc})
	[F_{\rm D}] {\cal S}_{11}\notag\\&+{\cal S}_{11}^\dag f_{\rm D}^{(+)}(\tilde{\varepsilon}_0+{L}_{\rm osc})[F^\dag_{\rm D}]\rho_{00}e^{x/\lambda}\Big\},\label{I2}
\end{align}
\esube
are the forward current from the QD out to the drain and the backward current from the drain into the QD, respectively. They are derived from the $\chi$-dependent quantum master equation (\ref{QME4}).
The results for the total stationary current $\lla I\rra$, its forward and backward components ($\lla I_+\rra$ and $\lla I_-\rra$), as well as the noise (in terms of the Fano factor $\lla I^2 \rra/\lla I\rra$) versus electric field ${\cal E}$ with a given bias voltage $V=5\omega_0$ are plotted in \Fig{fig4}(a)-(d), respectively, for various dampings.

\begin{figure}
\centering
\includegraphics[width=8.5cm]{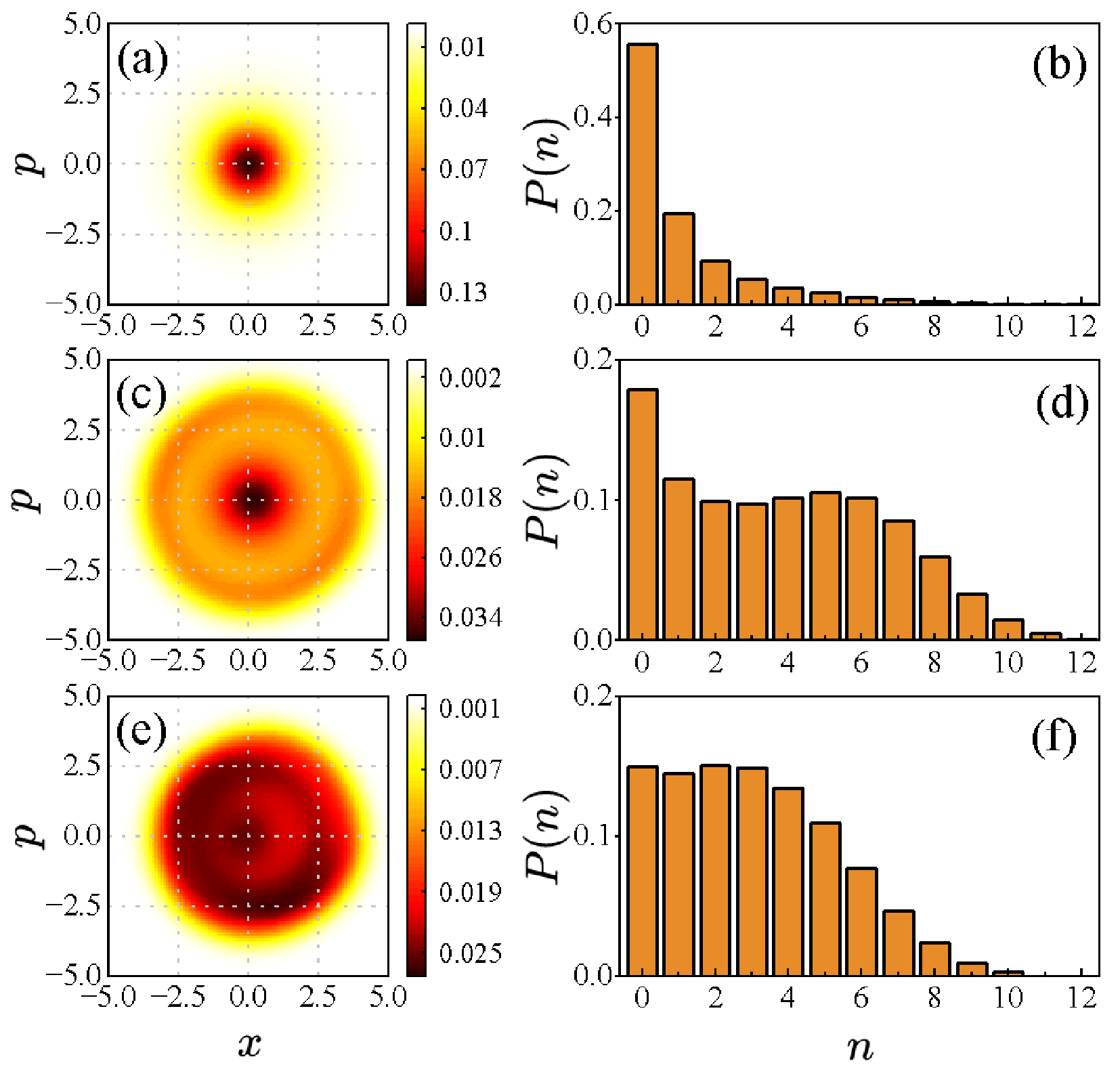}
\caption{\label{fig5}(a) The Wigner function distribution and (b) the phonon occupation probability distribution under the bias $V=5\omega_0$ for a weak electric field ${\cal E}=0.05 \omega_0/x_0$ (corresponding to $d=100x_0)$. (c) and (d) are the corresponding results for ${\cal E}=0.2 \omega_0/x_0$ ($d=25x_0$); (e) and (f) are the corresponding results for ${\cal E}=0.5 \omega_0/x_0$ ($d=10x_0$). The other plotting parameters are the same as those in \Fig{fig2}.}
\end{figure}

Let us first consider the situation of a weak damping ($\gam=0.02\omega_0$). As the electric field increases from ${\cal E}=0$ to ${\cal E}\approx 0.25 \omega_0/x_0$, more vibrational modes are activated and the oscillation amplitude increases. This leads to enhanced forward and backward currents, as seen in \Figs{fig4}(b) and (c). Since the forward current dominates, one observes an increasing total current as shown in \Fig{fig4}(a).
The Fano factor, however, is strikingly different: It shows a turnover behavior. It first rises and  reaches its maximum value roughly at ${\cal E}\approx 0.1\omega_0/x_0$ and then falls off with ${\cal E}$, see \Fig{fig4}(d). 
We ascribe this turnover behavior to the transition from the random tunneling regime to the coexistence regime where both stochastic tunneling and high-energy oscillatory tunneling exist. 
For instance, at ${\cal E}=0.05 \omega_0/x_0$ (corresponding to an effective distance $d=100x_0$) the electric field is not strong enough such that electrons can only be transported through the lowest vibrational modes, see the Wigner distribution and phonon occupation probability in \Fig{fig5}(a)-(b).
As ${\cal E}$ increases to $0.2 \omega_0/x_0$ (corresponding to an effective distance $d=25x_0$), high-energy phonon states are activated and may even dominate, see \Fig{fig5}(c)-(d).
In the phase space, one finds a ring progressively evolving out of the central spot, a signature indicating at least partially electron shuttling behavior.

As the electric field rises from ${\cal E}\approx 0.25  \omega_0/x_0$ to ${\cal E}\approx 0.3 \omega_0/x_0$, the backward current increases more rapidly than the forward one [see \Fig{fig4}(b) and (c)], which has a twofold effect on transport. 
First, due to the Coulomb blockade the oscillator cannot accommodate two electrons simultaneously.
Once a backward tunneling electron occupies the QD, tunneling of a second electron from the left electrode to the QD is not allowed, which suppresses the forward tunneling events.
Second, the rapid increase in the backward current itself leads to an inhibition of the total current [see \Eq{I}].
As a result, the total current reaches its maximum value at ${\cal E}\approx 0.25\omega_0/x_0$ and then decreases with ${\cal E}$ as shown in \Fig{fig4}(a).

As the electric field further increases from ${\cal E}\approx0.3\omega_0/x_0$ onwards, 
the backward current approaches gradually its maximum at roughly ${\cal E}\approx 0.45 \omega_0/x_0$ and then decreases with  ${\cal E}$. 
This is due to the fact that the backward moving electrons have to overcome not only the mechanical damping but also a stronger electric force. 
The system is inclined to work in the stochastic tunneling regime, which is indicated by the slightly decreased ring of the Wigner distribution as shown in \Fig{fig5}(e) for $d=10x_0$.
As a result, the noise tends gradually to the Poissonian value $\lla I^2\rra/\lla I\rra \to 1$, as shown in \Fig{fig4}(d).
Furthermore, we observe a slight suppression of the Wigner distribution on the upper right corner of the inner ring [see \Fig{fig5}(e)]. This reveals that at a large electric field the oscillator tends to drop the electron to the drain prior to achieving the maximum 
displacement on the right.

\begin{figure}
\centering
\includegraphics[width=8.5cm]{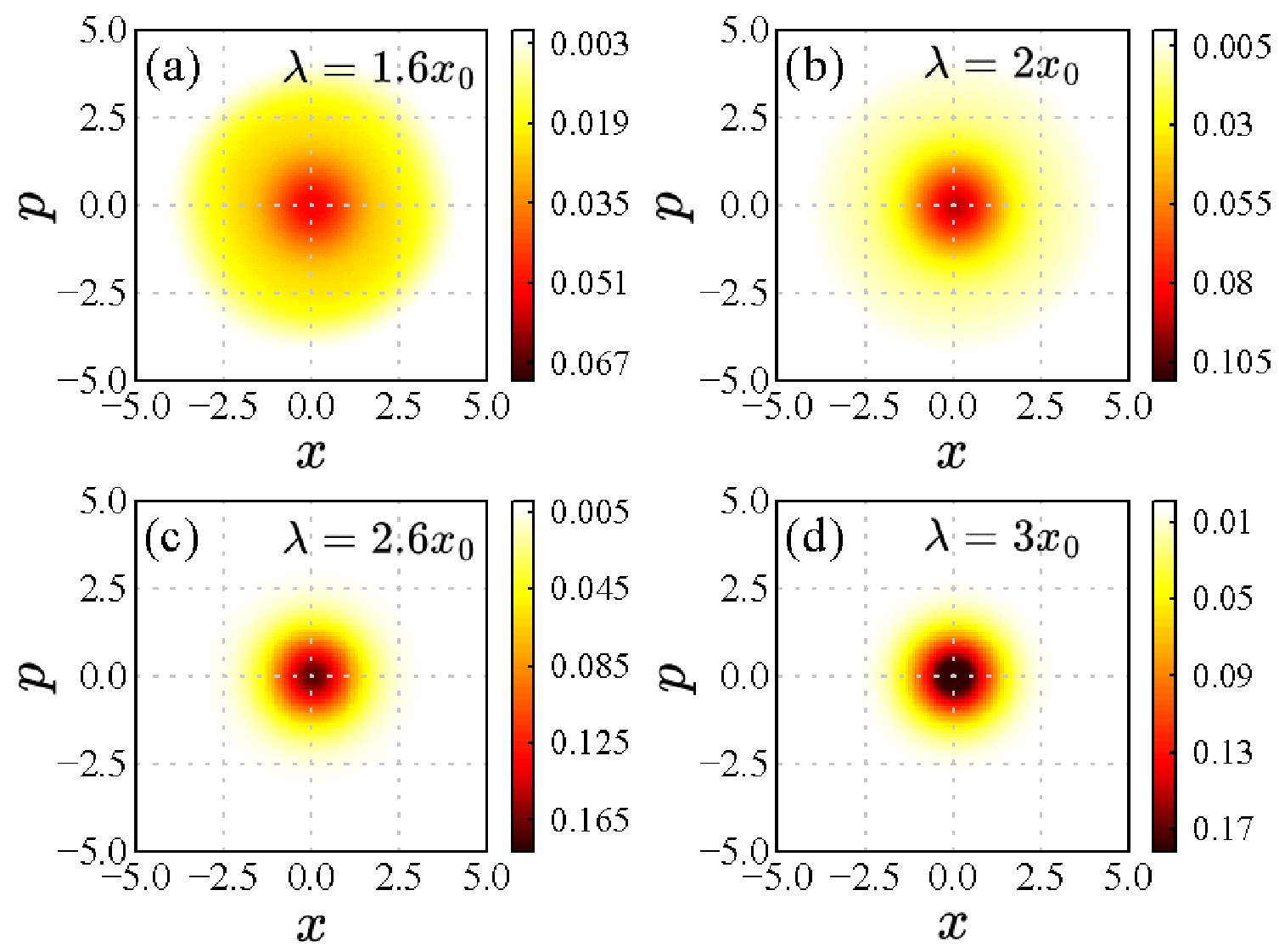}
\caption{\label{fig6} The Wigner distributions for a weak electric field ${\cal E}=0.1 \omega_0/x_0$ with different tunneling lengths (a) $\lambda=1.6x_0$, (b) $\lambda=2.0x_0$, (c) $\lambda=2.6x_0$, and (d) $\lambda=3.0x_0$, respectively. The other plotting parameters are the same as those in \Fig{fig2}.}
\end{figure}

\begin{figure}
	\centering
	\includegraphics[width=8.5cm]{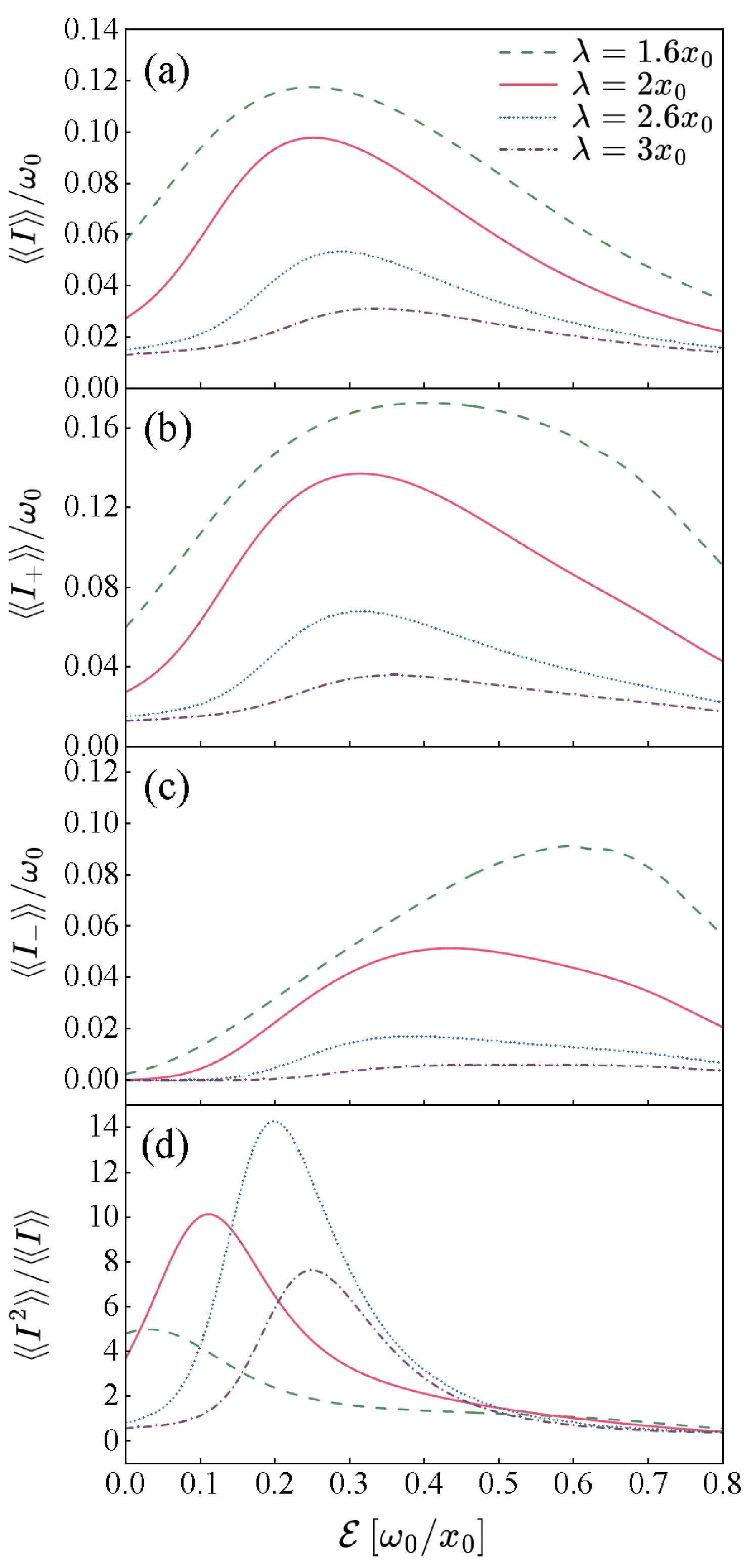}
	\caption{\label{fig7}(a) The stationary current ($\lla I\rra$), (b) forward ($\lla I_+\rra$), and (c) backward ($\lla I_-\rra$) components, and (d) the Fano factor ($\lla I^2\rra/\lla I\rra$) versus electric field under a given bias $V=5\omega_0$ for various tunneling lengths. The other plotting parameters are the same as those in \Fig{fig2}.}
\end{figure}

It is also instructive to investigate the influence of the tunneling length ($\lmd$) on transport characteristics.
We find that it introduces two effects. 
First, a growing $\lambda$ leads to a  quantum-to-classical crossover of the oscillator \cite{Nov03256801}. Furthermore, the involving classical fluctuations decrease because the oscillator is less sensitive to the external electric field.
This is indicated by a reduced size of the spot in the Wigner distribution, as shown in \Fig{fig6}(a)-(d) for a small electric field (${\cal E}=0.1 \omega_0/x_0$).
That means the larger $\lambda$, the more classical is the behavior of the harmonic oscillator and the less susceptible is the harmonic oscillator to the electric field.
Second, at an initially small tunneling length $\lambda$, further increasing it exponentially suppresses the effective tunneling rates, as implied in the tunneling Hamiltonian [\Eq{Htun}].
As a result, one observes a general inhibition of the forward, backward, and the total currents with increasing $\lambda$, cf. \Fig{fig7}(a)-(c).
Furthermore, for a smaller tunneling length, it requires a weaker electric field to excite the oscillator to higher vibrational modes in the shuttling regime. The peaks in the total current and noise are thus shifted toward smaller electric fields as $\lambda$ decreases, as shown in \Fig{fig7}(a) and (d), respectively.
On the contrary, the peaks in the backward current are shifted towards larger electric fields, cf. \Fig{fig7}(c). It is ascribed to the fact that  by reducing the tunneling length ($\lambda$) an electron can arrive at the source (backward motion) even for a stronger electric field.

\begin{figure}
\centering
\includegraphics[width=8.5cm]{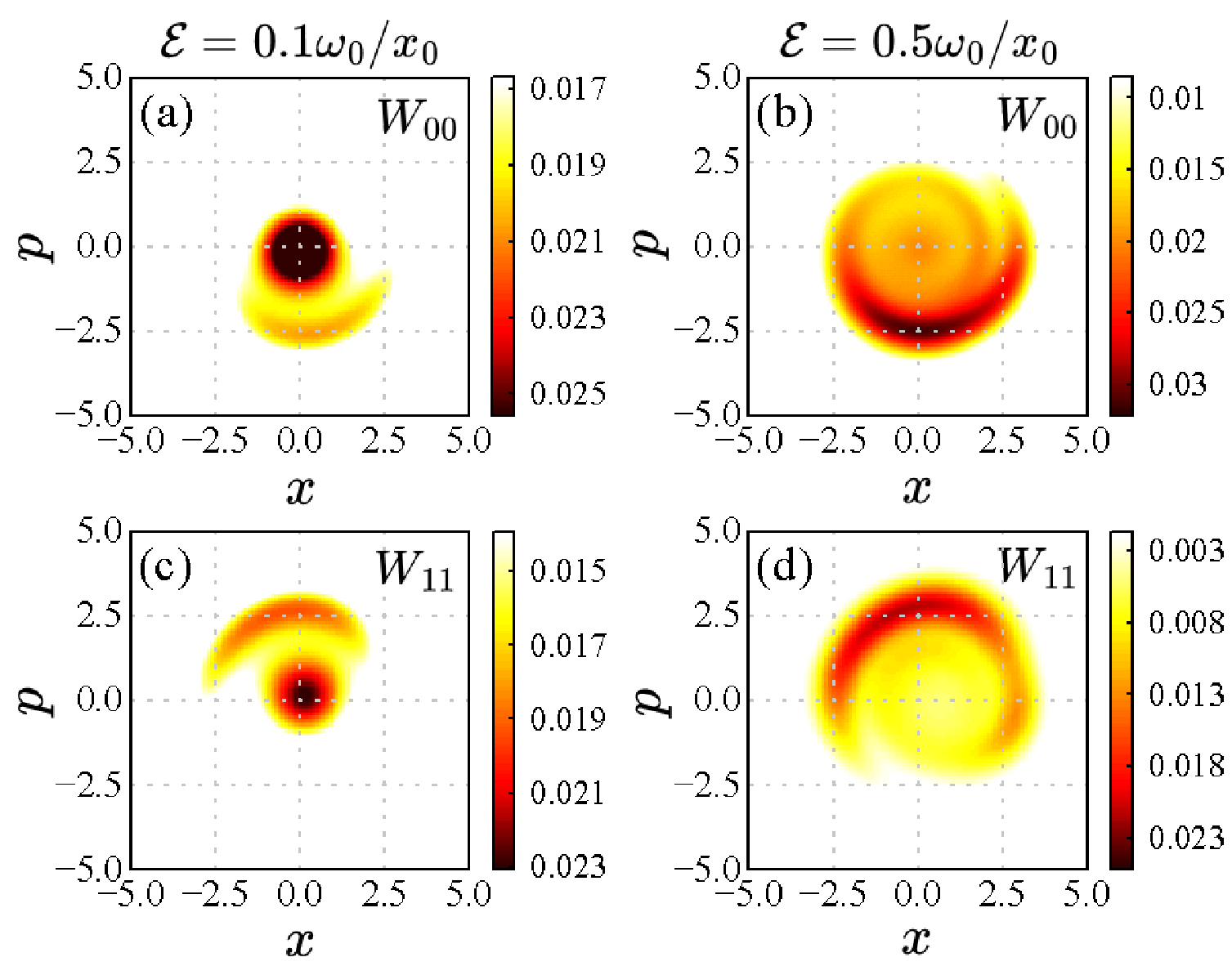}
\caption{\label{fig8} The charge-resolved Wigner distributions (a) $W_{00}$ and (c) $W_{11}$ for a weak electric field ${\cal E}=0.1 \omega_0/x_0$, and (b) $W_{00}$ and (d) $W_{11}$ for a strong electric field ${\cal E}=0.5 \omega_0/x_0$ with a small tunneling length $\lambda=1.6x_0$. The other plotting parameters are the same as those in \Fig{fig2}.}
\end{figure}

Strikingly, the strong backward tunneling at small $\lambda$ is uniquely reflected in the charge-resolved Wigner distributions $W_{00}(x,p)$ and $W_{11}(x,p)$, as shown in \Fig{fig8}.
For a weak electric field (${\cal E}=0.1 \omega_0/x_0$), the forward current dominates, see \Fig{fig7}(b). The oscillator shuttles the electron on its way from the source to the drain [a banana-shaped $W_{11}$ on the upper left corner in \Fig{fig8}(c)] and returns empty back [a banana-shaped $W_{00}$ on the lower right corner in \Fig{fig8}(a)]. 
In the presence of a strong electric field (${\cal E}=0.5 \omega_0/x_0$), we observe finite banana-shaped $W_{11}$ on the lower right corner and finite $W_{00}$
on the upper left corner [see \Fig{fig8}(b) and (d)], indicating tunneling of the electron from the drain to the source and then an empty oscillation after unloading the electron in the source.

So far, we have analyzed the transport characteristics as a function of the electric field at a weak damping. 
As the mechanical damping increases, the oscillator dissipates its energy to the environment and relaxs to the ground state more rapidly. It reduces the probability of triggering periodic electronic shuttlings via excited vibrational modes. 
Thus the forward, backward, total currents, as well as the noise are all suppressed compared to the situation of weak damping. 
Furthermore, the peaks are all shifted towards the direction of a stronger electric field, as it requires more energy to 
excite the oscillator in the case of a large damping.
In the limit of very strong damping ($\gam\gg \Gam_{\rm S}+\Gam_{\rm D}$), the oscillator gets quickly equilibrated
between electron tunneling events.
The dynamics would dominantly show two mechanically frozen charge states, i.e., the empty dot at rest in the oscillator's equilibrium position $|g\ra\otimes|x_{\rm eq}=0\ra$ and the charged dot in the shifted equilibrium position $|e\ra\otimes|x_{\rm eq}=\frac{e{\cal E}}{m\omg_0^2}\ra$. The electronic 
and mechanical states are approximately decoupled as: $\rho_{00}(t)=\sigma_{00}(t)\rho_{\rm osc}(g)$ and  $\rho_{11}(t)=\sigma_{11}(t)\rho_{\rm osc}(e)$, where $\rho_{\rm osc}(\xi)=\frac{e^{-\beta \la \xi|H_{\rmS}|\xi\ra}}
{{\rm tr}_{\rm osc}[e^{-\beta \la \xi|H_{\rmS}|\xi\ra}]}$ is
the canonical state of the uncharged ($\xi=g$) and charged ($\xi=e$) harmonic oscillator.
The quantum master equation (\ref{QME4}) effectively reduces to  
\bsube
\begin{align}
\dot{\sigma}_{00}(t)=-\tilde{\Gamma}_{\rmS}\sigma_{00}(t)+\tilde{\Gamma}_{\rmD}\sigma_{11}(t),
	\\
\dot{\sigma}_{11}(t)=-\tilde{\Gamma}_{\rmD}\sigma_{11}(t)+\tilde{\Gamma}_{\rmS}\sigma_{00}(t),
\end{align}
\esube
where the $\sigma_{00/11}$  are not matrices but c-numbers for the amplitudes of the empty and charged oscillator. It reproduces the equation for the two-state sequential tunneling
process \cite{Bla001,Gur9615932}, except for the difference of the renormalized tunneling rates
$\tilde{\Gamma}_{{\rm S/D}}=\Gamma_{{\rm S/D}}\tr_{\rm osc}[e^{\mp\frac{2x}{\lambda}}\rho_{\rm osc}(g/e)]$.
In this so-called ``renormalized tunneling'' regime, the stationary current and its noise are readily obtained as \cite{Nov04248302}
\bsube
\begin{align}
	\lla I\rra & =\frac{\tilde{\Gamma}_{\rmS}\tilde{\Gamma}_{\rmD}}{\tilde{\Gamma}_{\rmS}+\tilde{\Gamma}_{\rmD}},\label{cur1}
	\\
	\frac{\lla I^2\rra}{\lla I\rra} & =\frac{\tilde{\Gamma}^2_{\rmS}+\tilde{\Gamma}^2_{\rmD}}{(\tilde{\Gamma}_{\rmS}+\tilde{\Gamma}_{\rmD})^2}.\label{noise1}
\end{align}
\esube
The total stationary current $\lla I\rra = \lla I_+\rra$ and the noise are almost independent of the electric field, indicating that electron transfers through the lowest vibrational states and the electric field have a vanishing role to play. 
The corresponding Fano factor $\lla I^2\rra/\lla I\rra$ stays well below 1, as shown by the dotted curve in \Fig{fig4}(d).

\begin{figure}
\centering
\includegraphics[width=8.5cm]{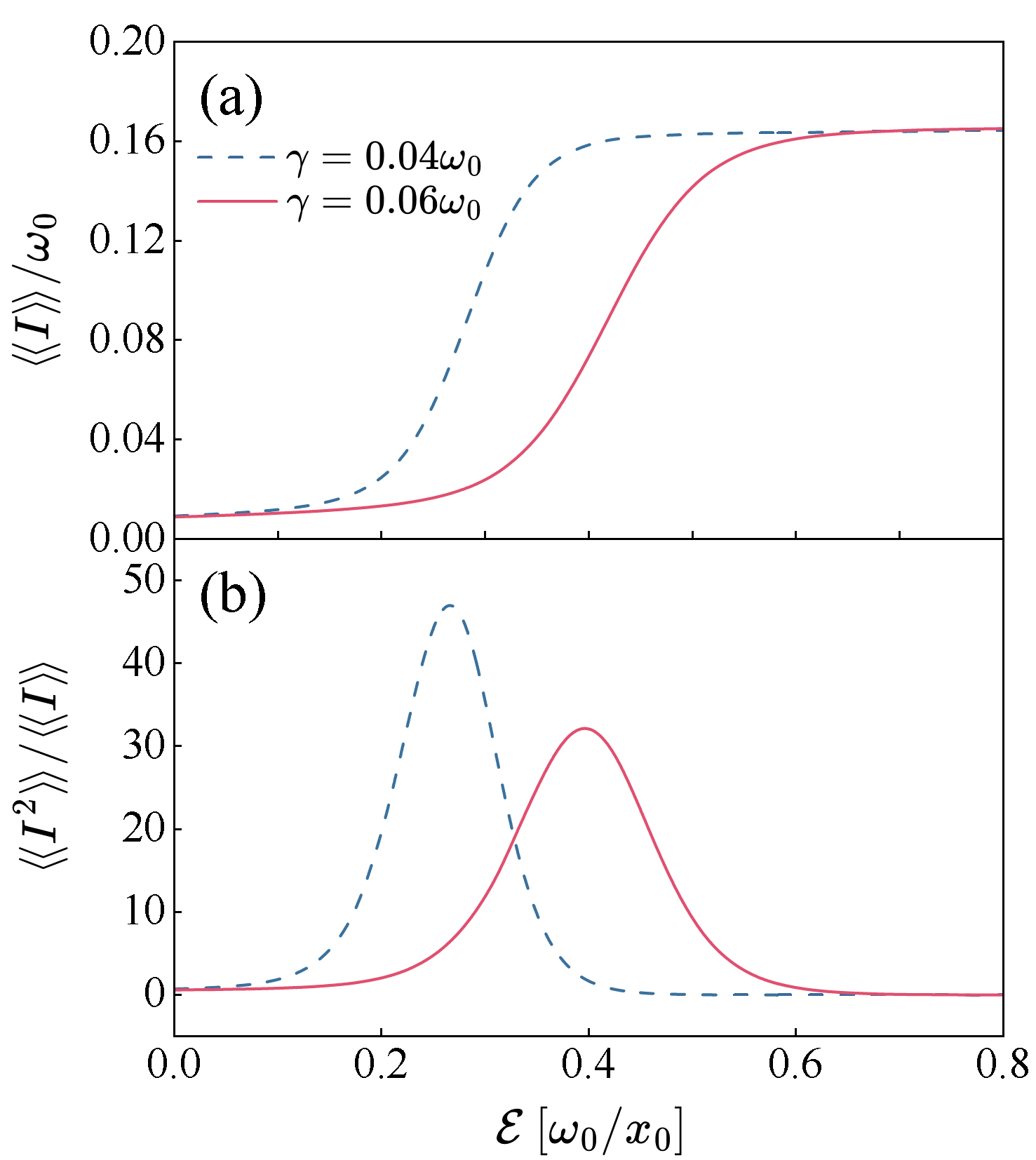}
\caption{\label{fig9}(a) Current and (b) the Fano factor versus electric field for different dampings in the limit of a large bias. The other plotting parameters are the same as those in \Fig{fig2}.}
\end{figure}

The quantum master equations developed here also allows us to investigate the transport properties at a large bias. Crucially, in the limit of large bias, \Eq{QME4} reduces to the quantum master equation widely used in the literature \cite{Nov03256801}.
The current versus the electric field is plotted in \Fig{fig9} for different dampings, where the bias is assumed to be the largest energy scale. 
It is observed that the current increases from a low (``renormalized tunneling'') value to a high (``shuttling'') one in a step-like way, where the position of the step depends on the damping rate, cf. \Fig{fig9}(a).
The corresponding Fano factor shows a remarkable peak at the position of the current step, as shown in \Fig{fig9}(b), indicating a strong coexistence of renormalized tunneling events and shuttling.
For a larger damping, it requires more electric energy to excite the oscillator such that the step in the current and peak in the noise are both shifted towards the direction of the larger electric field.
The peak in the Fano factor is relatively reduced for a larger damping, which implies a suppressed coexistence effect.
Anyway, in the limit of a large bias the current increases monotonically with the electric field. 
We can thus conclude that the anomalous decreasing current with rising electric field is solely a unique effect at low bias.

\section{CONCLUSIONS\label{thsec5}}

In this work, we have developed a quantum master equation for transport through a nanoelectromechanical device based on the second-order perturbation expansion in the system-environment coupling. 
This fully quantum approach is valid for arbitrary bias voltages, electro-mechanical couplings, and finite temperatures. 
We observed an anomalous current-electric field characteristics at small bias, i.e., the current decreases with a rising electric field
for a weak mechanical damping, in contrast to the situation of a large bias where the current never reduces with increasing electric field.  
Our investigation reveals that this unique feature is due to the fact that a larger electric field excites the oscillator to higher vibrational modes with a large oscillation amplitude, leading to an enhancement of backward electron tunneling events, as identified in both electron counting statistics and the phase space of charge-resolved Wigner functions.
Furthermore, the presence of a strong Coulomb interaction, for which double occupation is energetically prohibited, the rapid increase of backward tunneling events suppresses the forward current and eventually results in a suppressed current. 
In the opposite limit of strong damping, the oscillator quickly dissipates its energy to the environment and relaxs to the ground state. 
Electrons are thus dominantly transported through the lowest vibrational state and the electric 
field has a vanishing role to play. The net stationary current and the noise are almost 
independent of the electric field.
Our results demonstrate unambiguously the significance of the intriguing correlation between electronic and mechanical degrees of freedom in transport through a nanoelectromechanical 
device at low bias.

\begin{acknowledgments}
This work is supported by the National Natural Science Foundation of China (Grants No. 11774311, No. 12005188, and No. 12247158) and China CEEC Educational Exchange Program (Grant No. 2023269). 
S.R. acknowledges support from Ministry of Science, Technological Development and Innovation of the Republic of Serbia (Grants No. 451-03-137/2025-03/200125 and 451-03-136/2025-03/200125).
G.E. acknowledges the 
support by the National Natural Science Foundation of China (Grant No. W2432004). 
G.S. acknowledges financial support by the DFG through the CRC
1242 (Project-ID 278162697).
\end{acknowledgments}

\appendix

\section{\label{thAppA}DERIVATION OF THE QME}

Under the second-order Born-Markovian approximation in terms of the tunneling Hamiltonian \Eq{htun} in 
the displaced picture, the reduced density matrix satisfies \cite{Li05205304,Xu029196}
\begin{align}\label{APPAstt}
	\frac{d}{dt}\tilde{\varrho}(t)=&-\int_{0}^{\infty} d\tau \tr_\rmB[\tilde{\mathbf H}_{\rm tun}(t),[\tilde{\mathbf H}_{\rm tun}(t-\tau),\tilde{\varrho}(t)\otimes\rho_\rmB]]
	\nonumber \\
	=&-\int_{0}^{\infty} d\tau \{\tr_\rmB[\tilde{\mathbf H}_{\rm tun}(t)\tilde{\mathbf H}_{\rm tun}(t-\tau)\tilde{\varrho}(t)\otimes\rho_\rmB]
	\nonumber \\
	&-\tr_\rmB[\tilde{\mathbf H}_{\rm tun}(t)\tilde{\varrho}(t)\otimes\rho_\rmB \tilde{\mathbf H}_{\rm tun}(t-\tau)]
	\nonumber \\
	&-\tr_\rmB[\tilde{\mathbf H}_{\rm tun}(t-\tau)\tilde{\varrho}(t)\otimes\rho_\rmB \tilde{\mathbf H}_{\rm tun}(t)]
	\nonumber \\
	&+\tr_\rmB[\tilde{\varrho}(t)\otimes\rho_\rmB\tilde{\mathbf H}_{\rm tun}(t-\tau)\tilde{\mathbf H}_{\rm tun}(t)]\}
	\nonumber \\
	=&-[({\rm I})+({\rm II})+({\rm III})+({\rm IV})],
\end{align}
where $\rho_\rmB$ is the density matrix of the electrodes which are assumed to be in a local equilibrium,  $\tr_\rmB[(\cdots)]$ stands for the trace over the 
degrees of the freedom of the electrodes, and  $\tilde{\mathbf H}_{\rm tun}(t)=e^{i{\mathbf H}_0 t}\tilde{\mathbf H}_{\rm tun}e^{-i{\mathbf H}_0 t}$ is the tunneling 
Hamiltonian in the interaction picture defined in terms of the free Hamiltonian ${\mathbf H}_0={\mathbf H}_\rmS+{\mathbf H}_{\rm el}$.

There are in total four terms in \Eq{APPAstt}. Let us consider the calculation of the first term $({\rm I})$ as an example

\begin{widetext}
	\begin{align}
		({\rm I})=&\frac{1}{2}\sum_{\nu\nu'kk'}\int_{-\infty}^{\infty} d\tau t_{\nu k} t^\ast_{\nu'k'}
		\tr_\rmB\{
		[\tilde{F}_{\nu}(t) \tilde{c}^\dag_{\nu k}(t) \tilde{c}_0(t)][\tilde{c}_0^\dag(t-\tau)\tilde{c}_{\nu' k'}(t-\tau) 
		\tilde{F}^\dag_{\nu'}(t-\tau) ]\tilde{\varrho}(t)\otimes\rho_\rmB
		\nonumber \\
		&+[\tilde{c}_0^\dag(t)\tilde{c}_{\nu k}(t) \tilde{F}^\dag_{\nu}(t)][\tilde{F}_{\nu'}(t-\tau) \tilde{c}^\dag_{\nu' k'}(t-\tau) \tilde{c}_0(t-\tau)] \tilde{\varrho}(t)\otimes\rho_\rmB\},
	\end{align}
	where we have replaced $\int_0^\infty d\tau(\cdots)=\frac{1}{2}\int_{-\infty}^{\infty}d\tau(\dots)$ which means we have neglected the energy renormalization \cite{Bre07}, and the tilde is used to indicate a quantity in the interaction picture. The trace over the degrees of freedom of the electrodes gives rise to the correlation functions
	\begin{align}
		\tr_\rmB\{ \tilde{c}^\dag_{\nu k}(t) \tilde{c}_{\nu' k'}(t-\tau)\rho_\rmB\}&
		=f_\nu^{(+)}(\epsilon_{\nu k}) e^{i \epsilon_{\nu k} \tau} \delta_{\nu\nu'}\delta_{kk'},
		\\
		\tr_\rmB\{ \tilde{c}_{\nu k}(t)\tilde{c}^\dag_{\nu' k'}(t-\tau) \rho_\rmB\}
		&=f_\nu^{(-)}(\epsilon_{\nu k}) e^{-i \epsilon_{\nu k} \tau}\delta_{\nu\nu'}\delta_{kk'},
	\end{align}
	with $f_\nu^{(+)}(\epsilon_{\nu k})$ being the usual Fermi functions and $f_\nu^{(-)}(\epsilon_{\nu k}) =1-f_\nu^{(+)}(\epsilon_{\nu k})$.
	By transforming it back to the Schr\"{o}dinger picture $\varrho(t)=e^{-i {\mathbf H}_\rmS t} \tilde{\varrho}(t) e^{i {\mathbf H}_\rmS t}$, we arrive at 
	\begin{align}
		({\rm I})\to&\frac{1}{2}\sum_{\nu k}\!\int_{-\infty}^{\infty} \!\! d\tau |t_{\nu k}|^2
		\{\!
		F_{\nu} c_0 c_0^\dag e^{-i {H}_{\rm osc} \tau}\!F^\dag_{\nu}e^{i {H}_{\rm osc} \tau} \!f_\nu^{(+)}(\epsilon_{\nu k}) e^{i (\epsilon_{\nu k}-\tilde{\varepsilon}_0) \tau} \!\!+\!c_0^\dag F^\dag_{\nu}e^{-i {H}_{\rm osc} \tau}\! F_{\nu} e^{i {H}_{\rm osc} \tau}\! c_0  f_\nu^{(-)}(\epsilon_{\nu k}) e^{-i (\epsilon_{\nu k}-\tilde{\varepsilon}_0) \tau}\}\varrho(t)
		\nonumber \\
		=&\frac{1}{2} \sum_{\nu k}\! \int_{-\infty}^{\infty}\!\! d\tau |t_{\nu k}|^2
		\{
		F_{\nu} c_0 c_0^\dag e^{-i (L_{\rm osc}-\epsilon_{\nu k}+\tilde{\varepsilon}_0) \tau} [F^\dag_{\nu}] f_\nu^{(+)}(\epsilon_{\nu k}) 
		+c_0^\dag F^\dag_{\nu}e^{-i (L_{\rm osc}+\epsilon_{\nu k}-\tilde{\varepsilon}_0) \tau} [F_{\nu}]c_0f_\nu^{(-)}(\epsilon_{\nu k})\}\varrho(t)
		\nonumber \\
		=&\pi \sum_{\nu k} |t_{\nu k}|^2
		\{
		F_{\nu} c_0 c_0^\dag  \delta(L_{\rm osc}-\epsilon_{\nu k}+\tilde{\varepsilon}_0) [F^\dag_{\nu}] f_\nu^{(+)}(\epsilon_{\nu k})
		+c_0^\dag F^\dag_{\nu} \delta(L_{\rm osc}+\epsilon_{\nu k}-\tilde{\varepsilon}_0) [F_{\nu}]c_0f_\nu^{(-)}(\epsilon_{\nu k})\}\varrho(t),
	\end{align}
where we have introduced the superoperator associated with the free harmonic oscillator $L_{\rm osc}(\cdots)=[H_{\rm osc}, (\cdots)]$ satisfying $e^{-i L_{\rm osc} \tau}[\cdots]
=e^{-i{H}_{\rm osc}\tau} [\cdots] e^{i{H}_{\rm osc} \tau}$ with $H_{\rm osc}=\frac{p^2}{2 m}+\frac{1}{2}m \omg^2 x^2$ the Hamiltonian of the free harmonic oscillator.
	By introducing the tunneling width due to coupling between the electrode $\nu$ QD  $\Gamma_{\nu}(\omg)=2\pi\sum_{k}|t_{\nu k}|^2 \delta(\omg-\epsilon_{\nu k})$, the first term (I) simplifies to 
	\begin{align}
		({\rm I})\to&\frac{1}{2} \sum_{\nu} \{
		F_{\nu} c_0\Gam_{\nu}({L}_{\rm osc}+\tilde{\varepsilon}_0)f_\nu^{(+)}({L}_{\rm osc}+\tilde{\varepsilon}_0) [ c_0^\dag F^\dag_{\nu}]
		+c_0^\dag F^\dag_{\nu} \Gam_{\nu}(-{L}_{\rm osc}+\tilde{\varepsilon}_0)f_\nu^{(-)}(-{L}_{\rm osc}+\tilde{\varepsilon}_0)[F_{\nu}c_0]\}\varrho(t).
	\end{align}
\end{widetext}

In this work, we assume the wide-band limit such that the tunneling width is energy independent, i.e.,  $\Gam_{\nu}(\omg)=\Gam_{\nu}$. One thus has
\begin{align}
	({\rm I})\to \frac{1}{2}\sum_{\nu}& \Gam_{\nu}\{
	F_{\nu} c_0  f_\nu^{(+)}({L}_{\rm osc}+\tilde{\varepsilon}_0) [c_0^\dag F^\dag_{\nu}]
	\nonumber \\
	&+c_0^\dag F^\dag_{\nu} f_\nu^{(-)}(-{L}_{\rm osc}+\tilde{\varepsilon}_0)[F_{\nu}c_0]\}\varrho(t).
\end{align}

Finally, we transform it from the displaced picture back to the original basis via $\rho(t)={\cal S}^\dag \varrho(t){\cal S}$ and arrive at
\begin{align}
	({\rm I})\to&\frac{1}{2} \sum_{\nu}\! \Gam_{\nu}\{
	e^{\frac{x}{\lmd_{\nu}}} c_0  \Upsilon_{\nu}^{(+)}({L}_{\rm osc})\!+\!c_0^\dag e^{\frac{x}{\lmd_{\nu}}} 
	\Upsilon_{\nu}^{(-)}(-{L}_{\rm osc})\}\rho(t),
\end{align}
where we have introduced 
\bsube\label{UpsilonPM}
\begin{align}
	\Upsilon_{\nu}^{(\pm)}(+{L}_{\rm osc})&={\cal S}^\dag f_{\nu}^{(\pm)}(\tilde{\varepsilon}_0+{L}_{\rm osc})[c_0^\dag\!F_{\nu}^\dag]{\cal S},
	\\
	\Upsilon_{\nu}^{(\pm)}(-{L}_{\rm osc})&={\cal S}^\dag f_{\nu}^{(\pm)}(\tilde{\varepsilon}_0-{L}_{\rm osc})[F_{\nu}c_0] {\cal S}.
\end{align}
\esube

The other three terms (II), (III), and (IV) in \Eq{APPAstt} can be evaluated in an analogous manner. By collecting all these four terms and 
including the counting fields by following a standard procedure \cite{Naz03,Bag03085316,Esp091665}, one eventually arrives at \Eq{QME1}.

\section{\label{thAppB}Classical Limit}

To obtain the classical limit of the quantum master equation (\ref{QME}), we consider each term therein. 
The first term
\be
{\cal L}_{\rm coh}\rho(t)=-\rmi [H_{\rm S},\rho(t)]
\ee
describes the coherent evolution of the reduced system (QD-plus-oscillator) in \Eq{HS}.
In the classical limit, this term reduces to \cite{Str21180605}
\begin{align}\label{lcal}
{\cal L}_{\rm coh}\rho(t) \mapsto -v\frac{\partial}{\partial x}+\frac{\partial}{\partial v}\left[\frac{k}{m}x -\frac{e{\cal E}}{m}\xi\right] P_\xi (x,v;t),
\end{align}
where $P_\xi (x,v;t)$ is the probability density at time $t$ to find the oscillator at position $x$ and velocity $v$ 
for an empty ($\xi=0$) or occupied ($\xi=1$) oscillator.

The third term in \Eq{QME} accounts for the interaction of the oscillator with the heat bath, with its explicit form given by 
\Eq{caldamp}.
In the classical limit $\hbar\to0$ and making the replacement $\gam=\gam'/m$, it reduces to
\begin{align}\label{ldamp}
{\cal L}_{\rm damp}\rho(t) \mapsto \frac{\partial}{\partial v}\left[\frac{\gam'}{m}v+\frac{\gam'}{\beta_{\rm B} m^2}\frac{\partial}{\partial v}\right] P_\xi (x,v;t).
\end{align}
Equations (\ref{lcal}) and (\ref{ldamp}) together reproduce the generator for the dot occupation $\xi$ 
as shown in Ref. \cite{Str21180605}:
\begin{align}\label{lq}
	{L}_{\xi}=-v\frac{\partial}{\partial x}\!+\!\frac{\partial}{\partial v}\left[
	\frac{k}{m}x\!+\!\frac{\gam'}{m}v-\frac{e{\cal E}}{m}\xi+\frac{\gam'}{\beta_{\rm B} m^2}\frac{\partial}{\partial v}\right].
\end{align}

To obtain the classical limit of the second term in \Eq{QME}, we first expand the involved Fermi functions in \Eqs{UpsilonPM} as follows: 
\begin{align}
	f_{\nu}^{(\pm)}(\tilde{\varepsilon}_0\pm{L}_{\rm osc})\approx&  f_{\nu}^{(\pm)}(z)|_{z=\tilde{\varepsilon}_0}+ \partial_z f_{\nu}^{(\pm)}(z)|_{z=\tilde{\varepsilon}_0} (\pm{L}_{\rm osc})
	\nonumber \\
	&+\frac{1}{2}\partial^2_z f_{\nu}^{(\pm)}(z)|_{z=\tilde{\varepsilon}_0} (\pm{L}_{\rm osc})^2+\ldots
\end{align}
Equations (\ref{UpsilonPM}) then become
\bsube\label{TayExp1}
\begin{align}
	\Upsilon_{\nu n}^{(\pm)}(+{L}_{\rm osc})=&f_{\nu}^{(\pm)}(\tilde{\varepsilon}_0)[c_0^\dag e^{\frac{x}{\lmd_{\nu}}}]
	\nonumber \\
	&+\partial_z f_{\nu}^{(\pm)}(z)|_{z=\tilde{\varepsilon}_0} [\tilde{H}_{\rm osc},c_0^\dag e^{\frac{x}{\lmd_{\nu}}}]
	\nonumber \\
	&+\partial^2_z f_{\nu}^{(\pm)}(z)|_{z=\tilde{\varepsilon}_0}[\tilde{H}_{\rm osc}, [\tilde{H}_{\rm osc},c_0^\dag e^{\frac{x}{\lmd_{\nu}}}]]
	\nonumber \\
	&+ \cdots,
	\\
	\Upsilon_{\nu n}^{(\pm)}(-{L}_{\rm osc})=&f_{\nu}^{(\pm)}(\tilde{\varepsilon}_0)[e^{\frac{x}{\lmd_{\nu}}}c_0]
	\nonumber \\
	&-\partial_z f_{\nu}^{(\pm)}(z)|_{z=\tilde{\varepsilon}_0} [\tilde{H}_{\rm osc},e^{\frac{x}{\lmd_{\nu}}}c_0]
	\nonumber \\
	&+\partial^2_z f_{\nu}^{(\pm)}(z)|_{z=\tilde{\varepsilon}_0}[\tilde{H}_{\rm osc}, [\tilde{H}_{\rm osc},e^{\frac{x}{\lmd_{\nu}}}c_0]]
	\nonumber \\
	&+\cdots,
\end{align}
\esube	
where we have used ${\cal S}^\dag (c_0^\dag F_{\nu}^\dag){\cal S}=c_0^\dag e^{\frac{x}{\lmd_{\nu}}}$, ${\cal S}^\dag (F_{\nu} c_0) {\cal S}=e^{\frac{x}{\lmd_{\nu}}}c_0$, and
\begin{align}
	\tilde{H}_{\rm osc}=&{\cal S}^\dag H_{\rm osc} {\cal S}
	\nonumber \\
	=&\hbar\omega_0 \left(a^\dag a+\frac{1}{2}\right)-\left(e{\cal E}x -\frac{e^2 {\cal E}^2}{2m\omg_0^2}\right)c_0^\dag c_0.
\end{align} 
In the limit $\omg_0 \to 0$, one approximately has
\bsube\label{TayExp2}
\begin{align}
	[\tilde{H}_{\rm osc},c_0^\dag e^{\frac{x}{\lmd_{\nu}}}]\approx &-\left(e{\cal E}x -\frac{e^2 {\cal E}^2}{2m\omg_0^2}\right)c_0^\dag e^{\frac{x}{\lmd_{\nu}}},
	\nonumber \\
	[\tilde{H}_{\rm osc},[\tilde{H}_{\rm osc},c_0^\dag e^{\frac{x}{\lmd_{\nu}}}]]\approx &\left[-\left(e{\cal E}x -\frac{e^2 {\cal E}^2}{2m\omg_0^2}\right)\right]^2 c_0^\dag e^{\frac{x}{\lmd_{\nu}}},
	\\
	[\tilde{H}_{\rm osc},e^{\frac{x}{\lmd_{\nu}}}c_0]\approx & \left(e{\cal E}x -\frac{e^2 {\cal E}^2}{2m\omg_0^2}\right) e^{\frac{x}{\lmd_{\nu}}}c_0,
	\nonumber \\
	[\tilde{H}_{\rm osc}, [\tilde{H}_{\rm osc},e^{\frac{x}{\lmd_{\nu}}}c_0]] \approx & \left(e{\cal E}x -\frac{e^2 {\cal E}^2}{2m\omg_0^2}\right)^2 e^{\frac{x}{\lmd_{\nu}}}c_0.
\end{align}
\esube

By substituting \Eqs{TayExp2} into \Eqs{TayExp1}, we arrive at
\bsube\label{TayExp3}
\begin{align}
\Upsilon_{\nu}^{(\pm)}(+{L}_{\rm osc})\approx &f_{\nu}^{(\pm)}(\tilde{\varepsilon}_x) c_0^\dag e^{\frac{x}{\lmd_{\nu}}},
\\
\Upsilon_{\nu}^{(\pm)}(-{L}_{\rm osc})\approx &f_{\nu}^{(\pm)}(\tilde{\varepsilon}_x) e^{\frac{x}{\lmd_{\nu}}}c_0,
\end{align}
\esube	
where we have introduced $\tilde{\varepsilon}_x=\varepsilon_0-\frac{e^2 {\cal E}^2}{2m\omg_0^2}-e{\cal E}x$.
Inserting \Eqs{TayExp3} into \Eq{QME1}, we find (for $\chi=0$)
\begin{widetext}
\begin{align}
{\cal L}_{\rm tun}\rho(t)\!=\!
	-\!\!\sum_{\nu={\rm S,D}}\!\!\frac{\Gamma_\nu}{2}
	\!\left\{\!\left[ c_0e^{\frac{x}{\lambda_\nu}}\!, f_\nu^{(+)}(\tilde{\varepsilon}_x)c_0^\dag e^{\frac{x}{\lambda_\nu}}\rho(t)\!-\!\rho(t) 
	f_\nu^{(-)}(\tilde{\varepsilon}_x)c_0^\dag e^{\frac{x}{\lambda_\nu}}\right]	
\!+\!\left[c_0^\dag e^{\frac{x}{\lambda_\nu}}\!, f_\nu^{(-)}(\tilde{\varepsilon}_x)c_0 e^{\frac{x}{\lambda_\nu}}\rho(t)\!-\!\rho(t) 
	f_\nu^{(+)}(\tilde{\varepsilon}_x)c_0 e^{\frac{x}{\lambda_\nu}}\right]\!\right\}.
\end{align}
\end{widetext}

In the classical limit ($\hbar\to0$), the operator $x$ is mapped to a $c$-number such that
\begin{align}
{\cal L}_{\rm tun}\rho(t)\mapsto&
\sum_{\nu={\rm S,D}}\Gamma_\nu f_\nu^{(+)}(\tilde{\varepsilon}_x)e^{\frac{2x}{\lambda_\nu}}
{\cal D}[c_0^\dag]P(x,v;t)
\nonumber \\
&+\sum_{\nu={\rm S,D}}\Gamma_\nu f_\nu^{(-)}(\tilde{\varepsilon}_x)e^{\frac{2x}{\lambda_\nu}}
{\cal D}[c_0]P(x,v;t)
\nonumber \\
=& \sum_{\xi'} R_{\xi\xi'}(x)P_{\xi'}(x,v;t),\label{Rxi}
\end{align}
where ${\cal D}[A]\rho(t)=A\rho(t)A^\dag-\frac{1}{2}\{A^\dag A,\rho(t)\}$ is the Lindblad superoperator and $R_{\xi\xi'}(x)$ is implicitly defined in \Eq{Rxi}.
By collecting \Eqs{lq} and (\ref{Rxi}), one finally arrives at the classical limit of the quantum master equation (\ref{QME}) as shown 
in Refs. \cite{Str21180605,Wac19024001,Wac19073009}:
\begin{align}
\frac{\partial}{\partial t} P_\xi(x,v;t)=L_{\xi}P_\xi(x,v;t)+\sum_{\xi'} R_{\xi\xi'}(x)P_{\xi'}(x,v;t),
\end{align}
where $L_{\xi}$ is given in \Eq{lq}.


\end{document}